\documentclass[iop]{emulateapj}
\usepackage{amsmath}
\usepackage{rotating}

\def\stacksymbols #1#2#3#4{\def\theguybelow{#2}
        \def\verticalposition{\lower#3pt}
        \def\spacingwithinsymbol{\baselineskip0pt\lineskip#4pt}
        \mathrel{\mathpalette\intermediary#1}}
\def\intermediary #1#2{\verticalposition\vbox{\spacingwithinsymbol
        \everycr={}\tabskip0pt
        \halign{$\mathsurround0pt#1\hfil##\hfil$\crcr#2\crcr
                \theguybelow\crcr}}}

\shorttitle{THE DICHOTOMY IN COOLING-FLOW-FED ACCRETION OF SMBHs}
\shortauthors{GUO \& MATHEWS}

\begin{document}
\bibliographystyle{apj} 

\title {Hot versus Cold: the Dichotomy in Spherical Accretion of Cooling Flows onto Supermassive Black Holes in Elliptical Galaxies, Galaxy Groups and Clusters}

\author{Fulai Guo\altaffilmark{1,2,3} and William G. Mathews \altaffilmark{2}}

\altaffiltext{1}{ETH Z\"{u}rich, Institute for Astronomy, Wolfgang-Pauli-Strasse 27, CH-
8093, Z\"{u}rich, Switzerland; fulai.guo@phys.ethz.ch}

\altaffiltext{2}{UCO/Lick Observatory, Department of Astronomy and Astrophysics, University of California, Santa Cruz, CA 95064, USA}

\altaffiltext{3}{Zwicky Prize Fellow}

\begin{abstract}

Feedback heating from active galactic nuclei (AGNs) has been commonly invoked to suppress cooling flows predicted in hot gas in elliptical galaxies, galaxy groups and clusters. Previous studies have focused on if and how AGN feedback heats the gas, but little paid attention to its triggering mechanism. Using spherically symmetric simulations, we investigate how large-scale cooling flows are accreted by central supermassive black holes (SMBHs) in eight well-observed systems and find an interesting dichotomy. In massive clusters, the gas develops a central cooling catastrophe within about the cooling time (typically $\sim 100$ - $300$ Myr), resulting in a cold-mode accretion onto SMBHs. However, in our four simulated systems on group and galaxy scales at a low metallicity $Z=0.3Z_{\sun}$, the gas quickly settles into a long-term state which has a cuspy central temperature profile extending to several tens to about 100 pc. At the more realistic solar metallicity, two groups (with $R_{\rm e} \sim 4$ kpc) still host the long-term hot-mode accretion. Both accretion modes naturally appear in our idealized calculations where only cooling, gas inflow, and compressional heating are considered. The long-term hot-mode accretion is maintained by the quickly-established closeness between the timescales of these processes, preferably in systems with low gas densities, low gas metallicities, and importantly, compact central galaxies, which result in strong gravitational acceleration and compressional heating at the intermediate radii. Our calculations predict that central cuspy temperature profiles appear more often in smaller systems than galaxy clusters, which instead often host significant cold gas and star formation.
 
\end{abstract}


\section{Introduction}
\label{section:intro}

Massive elliptical galaxies, galaxy groups and clusters usually contain a large amount of hot diffuse gas with temperatures ranging from a few tenth of keV to about 10 keV. The hot gas emits prolifically in thermal X-rays, which have been frequently detected through  {\it ROSAT}, {\it Chandra}, and {\it XMM-Newton} observations (e.g., \citealt{peterson06}; \citealt{rp07}). The X-ray surface brightness of many systems strongly peaks at their centers, where gas densities are much higher than outer regions. The cooling time of the hot gas in central regions is often much shorter than the ages of their host systems, and at a distance of about $1$ kpc from their centers, the cooling time is often as short as $0.1$ Gyr (e.g., \citealt{sanderson06}). Radiative cooling reduces the gas entropy and induces cooling flows toward the center, most notably in massive clusters where the mass inflow rates are predicted to be as high as several hundred $M_{\sun}/$yr \citep{fabian94}. However, multi-wavelength data, in particular X-ray observations by {\it Chandra} and {\it XMM-Newton}, indicate that the predicted cooling flows are strongly suppressed. 

The most invoked mechanism to suppress cooling flows is feedback heating from central active galactic nuclei (AGNs; e.g., \citealt{bruggen02}; \citealt{ruszkowski02}). Radio and X-ray observations have shown that AGN jets created by supermassive black holes (SMBHs) at the centers of elliptical galaxies, galaxy groups and clusters interact strongly with their gaseous environments, producing radio bubbles, X-ray cavities, and weak shocks (see \citealt{mcnamara07} and \citealt{mcnamara12} for recent reviews). Previous studies have been focusing on answering two questions: (i) Can AGN feedback efficiently heat the hot gas, strongly suppressing cooling flows (e.g., \citealt{bruggen02}; \citealt{ruszkowski02}; \citealt{birzan04}; \citealt{vernaleo06}; \citealt{gaspari11b}; \citealt{gaspari11a}; \citealt{dubois11}; \citealt{martizzi12}; \citealt{Mendygral12}; \citealt{babul13})? (ii) how does AGN feedback heat the hot gas (e.g., \citealt{ruszkowski04}; \citealt{bruggen07}; \citealt{guo08a})? Here in this paper, we explore another key question: What triggers AGN events? In other words, we would like to investigate how SMBHs are fed in a typical AGN feedback loop. This question is much less investigated in the literature and our detailed study in this paper aims to provide a missing link between cooling flows and AGN feedback heating.

AGN events may be triggered as SMBHs accrete hot or cold gas from various origins \citep{best12}. The global stability analysis of \citet{guo08b} suggests that to stably maintain a galaxy cluster in the observed cool core state, the heating mechanism that offsets radiative cooling should be positively related to the cooling-induced mass inflow rate, i.e., the AGN heating rate should increase as the cooling flow becomes stronger. Otherwise, the cluster would evolve into a cooling catastrophe or a non-cool core state. This suggests that in a typical AGN feedback loop, AGN feedback may be directly triggered by the accretion of cooling flows onto the central SMBH.

The accretion of cooling flows by central SMBHs has been previously studied by \citet{quataert00}, who investigated spherically-symmetric steady-state models for the Virgo cluster. They found that, depending on the accretion rate, the cooling flow to accretion flow transition can be hot-mode (Bondi-like accretion near the SMBH with negative temperature gradients $dT/dr<0$) or cold mode (gas cooling off before reaching the SMBH), and argued that the hot-mode transition operates for most observed cooling flows. Similarly, \citet{mathews12} studied general steady-state transonic accretion flows when radiative cooling is present and also found that the flow behavior near the SMBH is bimodal, either hot-mode or cold-mode, depending on a dimensionless parameter composed of the black hole mass and flow properties.

Steady-state models, while physically intuitive, do not directly tell us which mode the cooling-flow-fed accretion sets in in a real system. Furthermore, the hot gas in real systems may not be in a steady state, but instead its properties may evolve with time due to periodic AGN feedback events. Numerical simulations may provide useful insights. However, previous simulations of cooling flows and AGN feedback usually focus on kpc and larger scales, while the SMBH accretion happens on parsec and smaller scales. Recently, \citet{li12} investigated the cooling flow to accretion flow transition in the Perseus cluster with hydrodynamical simulations, finding that a cooling catastrophe develops at the cluster center within about $300$ Myrs, roughly the initial central gas cooling time. This suggests that the cooling-flow-fed accretion onto SMBHs is cold-mode, as also found in simulations of \citet{gaspari13}, which further studies the potential roles of AGN heating and subsonic turbulence on the black hole accretion.

While progresses are being made in theory (though uncertainties on the roles of heating, turbulence, thermal conduction and stellar mass losses are still present), observations are yet to effectively explore the black hole accretion region. X-ray observations of cool cores of galaxy clusters often show gas temperature dropping toward the cluster center, suggesting that cooling does play a significant role. But for most clusters, observations could not resolve the central 1 kpc, and thus could not directly explore the black hole accretion flow. X-ray observations of some more nearby elliptical galaxies also resolved the region from $r\sim 0.1$ to $1$ kpc, revealing that the gas temperature profile is often quite flat in this region and in some systems (e.g., NGC 4374, NGC 4696, NGC 3115), the gas temperature increases toward the center (\citealt{allen06}; \citealt{humphrey06}; \citealt{werner12}; \citealt{wong11}). The flat temperature profiles and especially the central negative temperature gradients suggest the importance of heating processes, potentially from AGN feedback, or adiabatic heating due to gravitational infall, which may lead to the hot-mode SMBH accretion. By analyzing X-ray data of nine nearby elliptical galaxies, \citet{allen06} found a relation between the classic Bondi accretion rate \citep{bondi52} and the AGN jet power, suggesting the presence of the hot-mode Bondi accretion toward the SMBHs in these systems. However, this analysis depends on a likely-inaccurate extrapolation of the gas properties from the inner observational bins to the unresolved Bondi radius, and the more recent study by \citet{russell13} found a much weaker relation.

To explore the mystery of the cooling flow to accretion flow transition, we perform numerical simulations for a series of {\it real} observed systems spanning a large range in mass, ranging from massive galaxy clusters to smaller systems including galaxy groups and elliptical galaxies. In particular, we choose the observed gas temperature and density profiles as the initial conditions, and adopt the gravitational potential which includes the contributions from the dark matter halo, the central galaxy's stellar distribution and the central SMBH, and which is also based on current estimates from observations. The problem involves a very large dynamical range in radius, ranging from accretion flows on pc scales to cooling flows formed on scales up to about $100$ kpc (about 4 - 5 orders of magnitude), and thus 3-dimensional simulations are very expensive and time-consuming. In the paper, we focus on spherically-symmetric models, which allow us to investigate the problem in a large number of systems within a reasonable amount of time. 

The rest of the paper is organized as follows. In Section~\ref{section2}, we describe the details of our model, including basic equations and numerical setup. The results, presented in Section~\ref{section3}, show that even in an idealized spherically-symmetric inflow, the cooling gas can behave bimodally when reaching the central black hole accretion region. We study the possible origin of this dichotomy in Section \ref{section4} and discuss our results and observational prospects in Section \ref{section5}. We summarize our main conclusions in Section~\ref{section:conclusion}.
  
\section{The Model}
\label{section2}

\begin{table*}
 \centering
 \begin{minipage}{140mm}
  \renewcommand{\thefootnote}{\thempfootnote} 
  \caption{Parameters of the Gravitational Potential for the Systems}
  \begin{tabular}{@{}lcccccccc}
  \hline &$M_{\rm vir}$ &$M_{0}$&$r_{\rm s}$&$M_{*}$&$R_{\rm e}$&$M_{\rm BH}$&$R_{\rm inf}$& \\ Name&($M_{\sun}$)&($M_{\sun}$)&
        (kpc)&$(10^{11}M_{\sun}$)&(kpc)&($10^{9}M_{\sun}$)&(pc)&Reference   \\ \hline 
        Perseus ...............&$8.8\times 10^{14}$ &$5.3\times 10^{14}$&$481$&9.12& 15.3 &0.34&167&1, 2 \\ 
       Abell 1795  ..........&$7.1\times 10^{14}$ &$4.2\times 10^{14}$&$430$&5.35& 40.3 &1.66&862&1 \\ 
      Abell 2199  ..........&$6.6\times 10^{14}$ &$3.8\times 10^{14}$&$390$  & 13.1 &43.7&1.0&634&3, 4, 5, 6 \\ 
     Virgo  ..................&$4.2\times 10^{14}$ & $3.5\times 10^{14}$&$560$&4.97& 5.03 & 3.2&243&7, 1, 8  \\  
     NGC 4261  ..........& $6.7\times 10^{13}$ & $4.41\times 10^{13}$&$281.1$&2.64&3.4& 0.49&85&9, 10  \\ 
        NGC 4472  ..........& $3.3\times 10^{13}$ & $9.65\times 10^{12}$&$63.1$&1.63&4.0& 0.565&138&9, 11 \\      
         NGC 1407  ..........&$1.6\times 10^{13}$ &$4.0\times 10^{12}$&36.1&1.6&4.4 & 1.03&211&9, 12  \\          
        NGC 6482  .......... &$7.1\times 10^{12}$ & $1.78\times 10^{12}$&$27.8$&2.34&3.4& 0.562&97&9, 13 \\            
          \hline
\label{table1}
\end{tabular}
\\
Notes.-- For the definition of the parameters, please see Section \ref{subsect:potential}. The values of $M_{0}$ are derived from the virial mass $M_{\rm vir}$ (or the concentration $c$) and $r_{\rm s}$ given in the references. $R_{\rm inf}$ is the radius of influence of the SMBH, within which its gravitational acceleration dominates. \\ 
References.-- (1) \citet{chandran07}; (2) \citet{wilman05}; (3) \citet{zakamska03}; (4) \citet{dimatteo01}; (5) \citet{paturel03}; (6) \citet{graham96}; (7) \citet{mclaughlin98}; (8) \citet{macchetto97}; (9) \citet{humphrey06}; (10) \citet{ferrarese96}; (11) \citet{merritt01}; (12) \citet{spolaor08}; (13) \citet{panessa06}
\end{minipage}
\end{table*}

The primary goal of this paper is to study how cooling flows convert to black hole accretion flows in elliptical galaxies, galaxy groups and clusters. In particular, we focus on the difference of this flow transition in these systems with a large range in mass. We use hydrodynamic simulations to investigate the evolution of cooling flows formed on large scales (the whole $r \lesssim 100$ kpc region in massive clusters) and simultaneously, how the cooling gas behaves after entering the central region of the SMBH's gravitational influence. We focus on the competitive roles of radiative cooling and gravitational heating in a spherically symmetric inflow, while ignoring other more complicated physics, e.g., angular momentum, additional heating sources (e.g. AGN feedback), turbulence, transport processes (thermal conduction and viscosity), and stellar mass losses from the central galaxy. The importance of these additional physics is less certain, but some of them may play a significant or even dorminant role in some systems (\citealt{brighenti02}; \citealt{gaspari12}; \citealt{li12}; \citealt{gaspari13}). Our calculations are a step forward from steady state models (\citealt{bondi52}; \citealt{quataert00}; \citealt{mathews12}), and may provide important insights when comparing with other works incorporating additional physics.

Unimpeded, long-lasting cooling flows are commonly considered to be ruled out by multi-wavelength observations of galaxy clusters \citep{peterson06}. AGN feedback heating is often considered to be responsible for the suppression or shut off of cooling flows. Observations of AGN bubbles indicate that AGN feedback events have duty cycles \citep{mcnamara07}, implying that AGN heating and the development of cooling flows are intermittent. It may be possible that between two successive generations of AGN feedback events, cooling flows could be established to some level, leading to the formation of cold gas and stars in the central galaxy as observed in many cool core clusters (\citealt{peterson06}; \citealt{odea08}). The development of cooling flows may also feed the central SMBH, triggering the next generation of AGN feedback events which shuts off cooling flows and completes the AGN feedback loop. Thus our calculations presented in this paper investigate how cooling flows develop and feed the central SMBH before the next generation of AGN feedback is triggered. This is obviously an idealized scenario, and the additional physical processes mentioned in the previous paragraph may play important roles as well. Furthermore, local thermal instabilities have also been considered to produce cold gas in cool core clusters (\citealt{mccourt12}; \citealt{sharma12}; \citealt{gaspari12b}), though \citet{li12} showed that local thermal instabilities do not occur before the central cooling catastrophe when AGN heating and turbulence are ignored.

With radiative cooling, the hydrodynamic evolution of thermal gas can be described by the following three equations:

\begin{eqnarray}
\frac{d \rho}{d t} + \rho \nabla \cdot {\bf v} = 0,\label{hydro1}
\end{eqnarray}
\begin{eqnarray}
\rho \frac{d {\bf v}}{d t} = -\nabla P-\rho \nabla \Phi ,\label{hydro2}
\end{eqnarray}
\begin{eqnarray}
\frac{\partial e}{\partial t} +\nabla \cdot(e{\bf v})=-P\nabla \cdot {\bf v}-n_{\text{e}}n_{\text{i}}\Lambda(T, Z) 
   \rm{ ,}\label{hydro3}
   \end{eqnarray}
\noindent
where $d/dt \equiv \partial/\partial t+{\bf v}  \cdot \nabla $ is the Lagrangian time derivative,
$\rho$ is the gas density, ${\bf v}$ is the gas velocity, $P=(\gamma-1)e$ is the gas pressure, $\gamma=5/3$ is the adiabatic index of thermal gas, $e$ is the gas energy density, $n_{\rm e}$ is the electron number density, and $n_{\rm i}$ is the ion number density.

The rightmost term in equation (\ref{hydro3}), $n_{\rm{e}}n_{\text{i}}\Lambda(T, Z)$, is the gas cooling rate.  In most simulations, we use the analytic cooling function in \citet{tozzi01}, which is based on calculations by \citet{sd93},
\begin{eqnarray}
  n_{\rm{e}}n_{\text{i}}\Lambda(T, Z)  &=1.0 \times 10^{-22} \left(\frac{n_{\rm{i}}}{\rm{cm}^{-3}}\right) \left(\frac{n_{\rm{e}}}{\rm{cm}^{-3}}\right)\;\;\;\;\;\;\;\;\;\;\;\;\;\;\;\; \nonumber \\
  &\times  \left[C_{1}\left(\frac{k_{\rm{B}}T}{\rm{keV}}\right)^{\delta_{1}}+C_{2}\left(\frac{k_{\rm{B}}T}{\rm{keV}}\right)^{\delta_{2}}+C_{3}\right]  \; \frac{\rm{erg}}{\rm{cm}^{3}\; {\rm s}}\rm{ .} \label{hydro21}
\end{eqnarray}
For an average metallicity $Z=0.3Z_{\sun}$, the constants are $\delta_{1}=-1.7$, $\delta_{2}=0.5$, $C_{1}=8.6\times 10^{-3}$, $C_{2}=5.8\times 10^{-2}$ and $C_{3}=6.3\times 10^{-2}$, and we can approximate $n_{\rm{i}}n_{\rm{e}}=0.704(\rho/m_{\rm{p}})^{2}$, where $m_{\rm{p}}$ is the proton mass. In Section 3.4, we presented a different set of simulations with the \citet{sd93} cooling function at $Z=Z_{\sun}$, and studied the dependence of our results on the gas metallicity. We manually truncate the cooling below a minimum temperature of $0.03$ keV. We are not interested in the evolution of cold gas in this paper, but instead we focus on studying {\it whether or not} the hot gas cools off, forming a cooling catastrophe as cooling flows develop.

According to the ideal gas law, the gas pressure is related to the gas temperature $T$ and the electron number density $n_{\rm e}$ via
\begin{eqnarray}
 P =\frac{\rho k_{B} T}{\mu m_{\mu}}=\frac{\mu _{e}}{\mu}n_{\rm e}k_{B}T , \label{estate}\\ \nonumber
\end{eqnarray}
\noindent
where $k_{B}$ is Boltzmann's constant, $m_{\mu}$ is the atomic mass unit, and $\mu =0.62 $ and $\mu_{e}=1.18$ are the mean molecular weight per thermal particle and per electron, respectively. 
We define the gas cooling time as
\begin{eqnarray}
t_{\rm cool}=\frac{e}{ n_{\rm{e}}n_{\text{i}}\Lambda(T) } \;{\rm ,}\label{tcooldef}
  \end{eqnarray}
and the gas entropy $S$ as (ignoring constants and logarithms; e.g., \citealt{lpc00}) 
\begin{eqnarray}
S=\frac{k_{\rm B}T}{ n_{\rm{e}}^{2/3} } \;{\rm .} \label{sdefinition}
  \end{eqnarray}

\subsection{Gravitational Potential}
\label{subsect:potential}

The gravitational potential $\Phi$ is contributed by three static components:
\begin{eqnarray}
\Phi = \Phi_{\text{DM}} +  \Phi_ {*} + \Phi_{\text{BH}}  \text{,}  
\end{eqnarray}
\noindent
where $\Phi_{\text{DM}}$ is the contribution from the dark matter halo, $\Phi_ {*}$ is the contribution from the stellar mass of the central galaxy, and $\Phi_{\text{BH}}$ is the contribution from the central SMBH. We ignore the contribution from the hot gas, which is very small compared to the other three components. We take a Navarro-Frenk-White (NFW) profile \citep{navarro97} for the dark matter halo:
\begin{eqnarray}
\rho_{\text{DM}}(r)=\frac{M_{0}/2\pi}{r(r+r_{\text{s}})^{2}}\text{,}  
\end{eqnarray}
\noindent
where $r_{\text{s}}$ is  the standard scale radius  of the NFW profile and $M_{0}$ is a characteristic mass. As discussed in \citet{zakamska03}, $r_{\text{s}}$ and $M_{0}$ are related to the virial mass $M_{\rm vir}$ and viral radius $r_{\rm vir}$, which is defined in this paper as the radius within which the mean dark matter density is $200 \rho_{\rm crit}(z)$, where $\rho_{\rm crit}(z)=3H(z)^{2}/8\pi G$ is the critical density of the universe at the redshift of the system. The corresponding dark matter gravitational potential is:
\begin{eqnarray}
\Phi_{\text{DM}}= -\frac{2GM_{0}}{r_{\text{s}}} \frac{\text{ln}(1+r/r_{\text{s}})}{r/r_{\text{s}}}      \text{,}  
\end{eqnarray}
\noindent
where $G$ is the gravitational constant. 

We take the stellar mass density to have a Hernquist profile \citep{hernquist90}:
\begin{eqnarray}
\rho_ {*} = \frac{M_ {*}a}{2\pi r}\frac{1}{(r+a)^{3}}     \text{,}  
\end{eqnarray}
\noindent
where $M_ {*}$ is the total stellar mass and $a$ is a scale length equal to $R_{e}/1.8153$, where $R_{e}$ is the radius of the isophote enclosing half the galaxy's light. The corresponding gravitational potential is:
\begin{eqnarray}
\Phi_{*}= -\frac{GM_ {*}}{r+a}     \text{.} 
\end{eqnarray}

We take the SMBH's gravitational potential to be given by
\begin{eqnarray}
\Phi_{\rm BH}= -\frac{GM_{\rm BH}}{r-r_{\rm g}}     \text{,} 
\end{eqnarray}
\noindent
where $M_{\rm BH}$ is the mass of the central SMBH, $r_{\rm g}=2GM_{\rm BH}/c^{2}$ is the Schwarzschild radius, and the $1/(r-r_{\rm g})$ mimics the effects of general relativity (\citealt{paczynsky80}; \citealt{quataert00}).

In this paper, we studied the transition of cooling flows to accretion flows in eight well-observed systems, including four galaxy clusters ($M_{\rm vir}>10^{14}M_{\sun}$) and four smaller systems, of which three have group-scale halos and one has a galaxy-scale halo, demarcated at $M_{\rm vir}=10^{13}M_{\sun}$. The gas temperature and density profiles of these systems are well determined by X-ray observations, and serve as our initial conditions (see Sec. 2.3). The model parameters of the gravitational potential in all the systems are also constrained by observations and directly adopted from previous publications. Ordered by the decreasing virial mass of the dark matter halo, we list the model parameters of these systems in Table 1. The references where we adopt the model parameters are listed in the rightmost column of Table 1. In particular, we note that $M_ {*}$ and $R_{e}$ in most systems are taken from \citet{chandran07} or \citet{humphrey06}. For Abell 2199, we derive the value of $M_ {*}$ from its brightest cluster galaxy NGC 6166's B-band luminosity ($M_{B}=-23.01$ adopted from the Hyperleda database of \citealt{paturel03}) according to the method described in \citet{chandran07} and take the value of $R_{e}$ from \citet{graham96}. The eighth column in Table 1 shows $R_{\rm inf}$, the radius of influence of the central SMBH, within which its gravitational acceleration dominates over that from the other two components. 

We note that although a series of eight well-observed systems are studied in the paper, they are not intended to be representative of all galaxy groups and clusters. Our sample does not include systems with $R_{e}\sim 6-14$ kpc, and in particular, the four small systems in our sample have $R_{e}\sim 3.4-4.4$ kpc, which lies in the lower part of the $R_{e}$ range for galaxy groups and plays an important role in the transition of cooling flows to accretion flows as shown in Section 4.3. Approximating the stellar distribution of the central galaxy by the Hernquist profile is crude, especially for brightest cluster galaxies (BCGs), which often have a Sersic index of $n>4$ \citep{graham96}. Future studies with a better (but more complicated) modeling of the central galaxy will be useful in exploring its accurate role in the evolution of cooling flows, though the development of a central cooling catastrophe in galaxy clusters is usually thought to be general \citep{mcnamara07}. 

\subsection{Simulation Setup}

Assuming spherical symmetry, we use the ZEUS-3D hydrodynamic code \citep{stone92} in its one-dimensional mode. To study how large-scale cooling flows are converted into SMBH accretion flows, an extremely large range in radius needs to be covered in the simulations. Most of the simulations presented in the paper were performed with an inner boundary of $r_{\text{min}}=10$ pc, which is usually small enough for us to investigate the fate of large-scale cooling flows near SMBHs. We have also run a few additional simulations with a smaller inner boundary $1$ pc (presented in Sec. 3.3 and Appendix A), but the results usually do not change appreciably. We choose the outer boundary to be $r_{\text{max}}=200$ kpc, which is larger than the cooling radii of all systems studied in this paper. We have also experimented with a larger outer boundary at $2$ Mpc and the results do not change (the difference is often negligible; see Appendix A). We typically ran each simulation for $3$ Gyr, but if a cooling catastrophe develops within this duration, we stop the simulation shortly after the cooling catastrophe. 

In order to resolve adequately the inner regions, we adopt a logarithmically spaced grid in which $(\Delta r)_{i+1}/(\Delta r)_{i}=(r_{\text{max}}/r_{\text{min}})^{1/N}$, where $N$ is the number of active zones. The simulations presented in this paper were ran with $N=1000$, corresponding to a grid size of  about $0.1$ pc near the inner boundary. We have also run a few simulations with $N=4000$ and got the same results. The common zero-gradient boundary conditions are used to allow for inflow and outflow at the grid boundaries, and in addition, we further assume that the gas is in contact with a thermal bath of constant temperature and pressure at the outer boundary, where the cooling time exceeds the Hubble time. Thus, we ensure that, for each individual system, the temperature and density of thermal gas at the outer boundary are fixed to their initial values, which are derived or extrapolated from observations, as further explained in the following subsection.

\subsection{Initial Conditions}
 
To investigate the cooling-flow-fed accretion in {\it real} ellipticals, galaxy groups and clusters, we choose the initial conditions to mimic the observed temperature and density profiles. We also assume that the hot gas is initially in hydrostatic equilibrium. To this end, we first choose an analytic fit to the observed temperature profile for each system studied in this paper, and then solve the density profile from hydrostatic equilibrium and the gravitational potential presented in Section \ref{subsect:potential}. The normalization of the density profile (specifically $n_{\rm e}$ at $1$ kpc) is determined so that the overall density profile fits the observed density profile reasonably well.

The temperature profiles of some systems differ significantly from those of others, and thus we could not use one ``universal" profile for all the systems (as also discussed in \citealt{humphrey06}). For the Perseus cluster, we took the analytic temperature fit to the observational data from \citet{churazov04}. For the cluster Abell 1795, we adopt the analytic temperature profile from \citet{guo10}, which fits very well the observational data from a few kpc to about $1000$ kpc. For the other six systems, we use one of the following analytic fits:
\begin{eqnarray}
T(r)=& T_{0} &- (T_{0}-T_{1})e^{-r/(2r_{c1})}    \text{,} \label{fit1} \\
T(r)=& T_{0} &- T_{1}e^{-r^{2}/(2r^{2}_{c1})}    \text{,} \label{fit2}\\
T(r)=& T_{0} &+T_{1}e^{-r/(2r_{c1})}    \text{,} \label{fit3}\\
T(r)=&T_{0} &- T_{0}e^{-r^{2}/(2r^{2}_{c1})} +T_{1} e^{-r^{3}/r^{3}_{c2}} \nonumber \\
       &+&T_{2}e^{-(r/r_{c3})^{1.3}} \text{.}\label{fit4}
\end{eqnarray}
For the cluster Abell 2199, we adopted equation (\ref{fit1}), with $T_{0}=4.3$ keV, $T_{1}=1.4$ keV, and $r_{c1}=13$ kpc. For the Virgo cluster, we followed \citet{ghizzardi04}, using equation (\ref{fit2}), with $T_{0}=2.399$ keV, $T_{1}=0.776$ keV, and $r_{c1}=18.15$ kpc. For the group NGC 4261, we adopted equation (\ref{fit1}), with $T_{0}=1.3$ keV, $T_{1}=0.6$ keV, and $r_{c1}=8$ kpc. For the elliptical galaxy NGC 4472, which is located in the Virgo cluster, we adopted equation (\ref{fit1}), with $T_{0}=1.17$ keV, $T_{1}=0.6$ keV, and $r_{c1}=5$ kpc. For the group NGC 1407, we adopted equation (\ref{fit4}), with $T_{0}=0.8$ keV, $T_{1}=T_{0}/3.8$, $T_{2}=T_{0}/1.3$, $r_{c1}=6.5$ kpc, $r_{c2}=1$ kpc, and $r_{c3}=50$ kpc. For the group NGC 6482, we used equation (\ref{fit3}), with $T_{0}=T_{1}=0.4$ keV, and $r_{c1}=8$ kpc. Together with observational data, the initial temperature and density profiles for each system are shown as solid lines in Figures 1, 3, 4, and 5. Our analytic temperature profiles and the derived initial gas density profiles fit with observational data reasonably well in all the systems presented in the paper.

 \begin{figure*}
   \centering
\plottwo {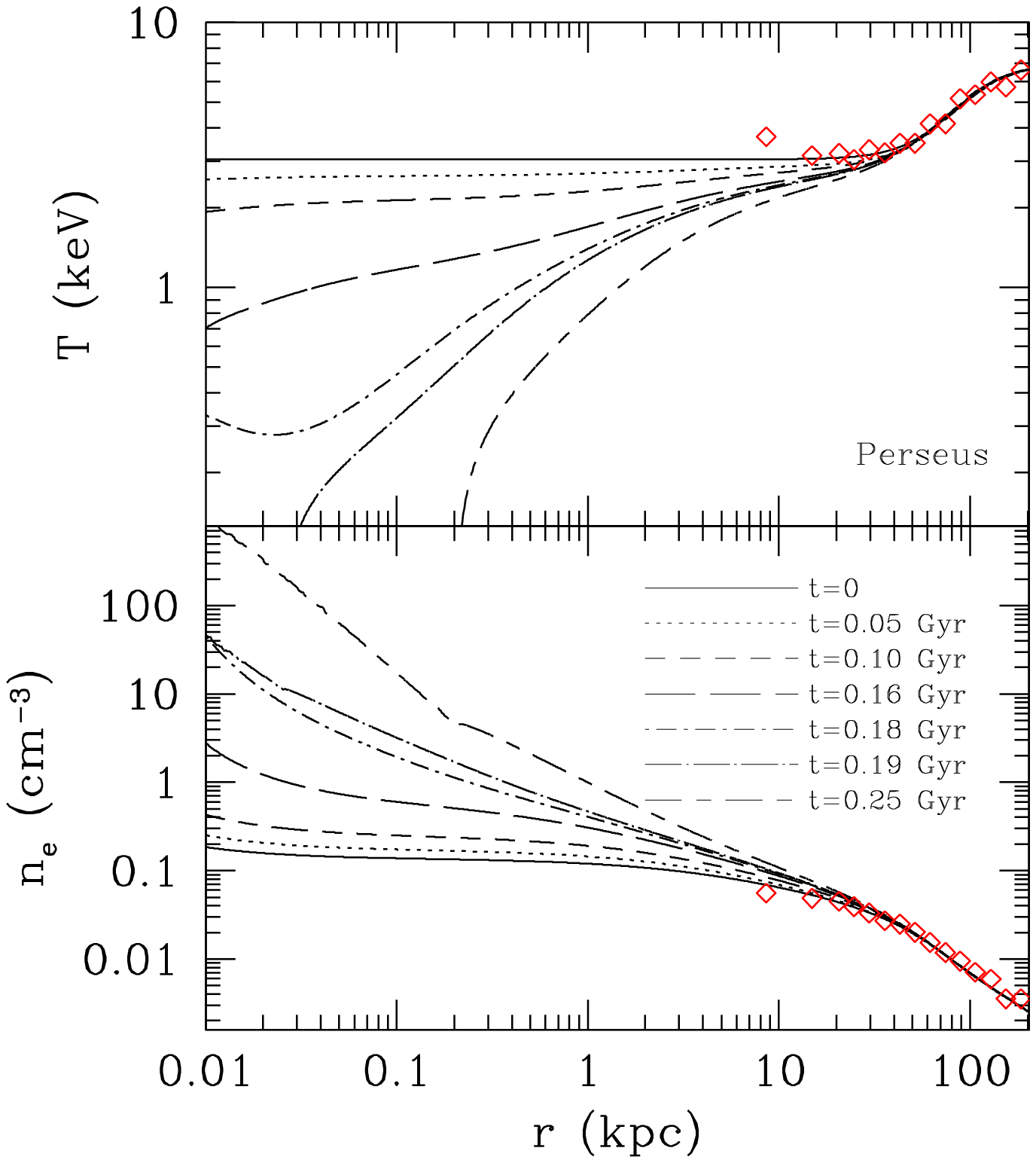} {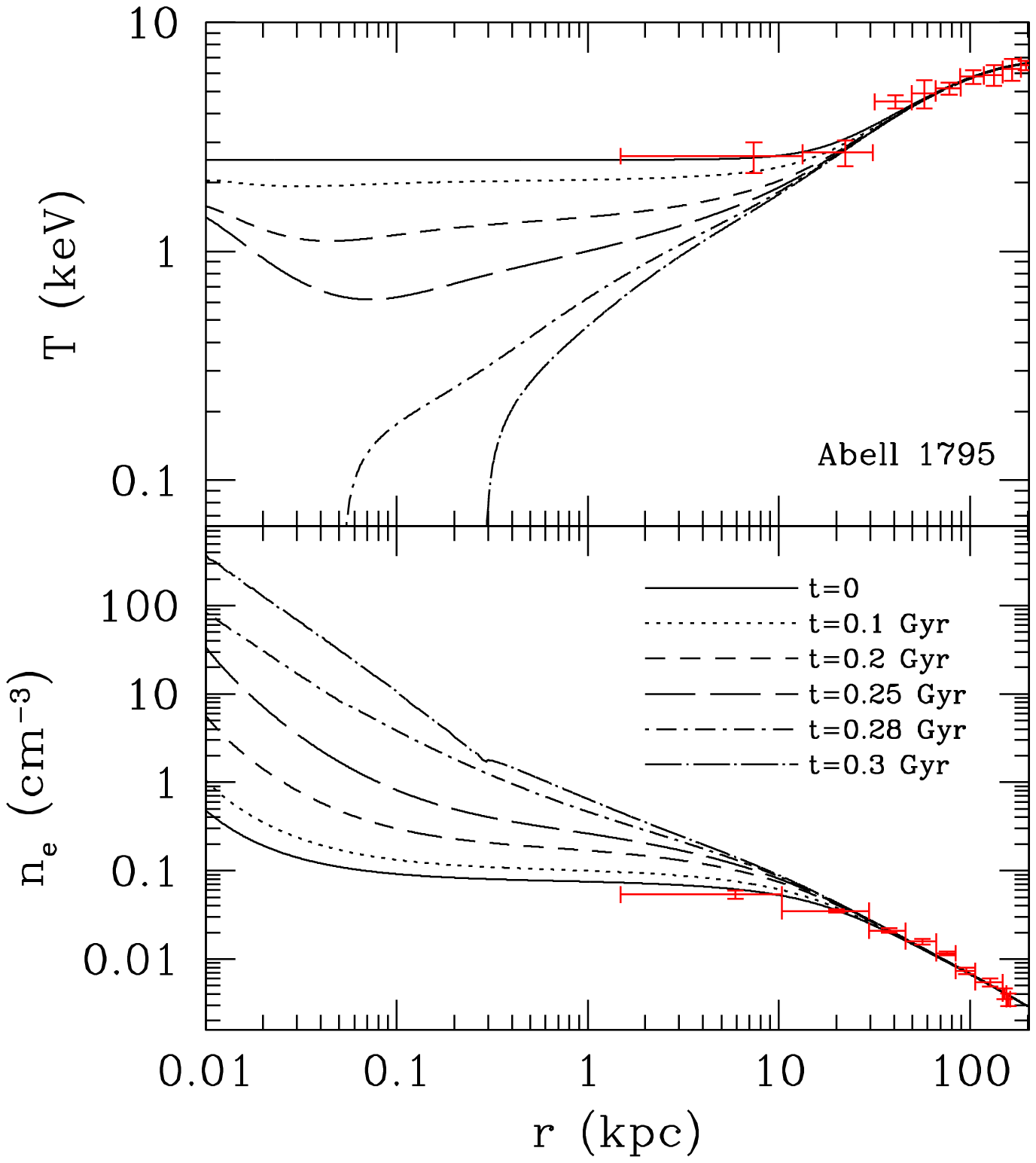}
\caption{The temporal evolution of temperature and electron number density profiles for the Perseus cluster (left) and Abell 1795 (right). The diamonds in the left panel correspond to XMM-Newton data of \citet{churazov03}, rescaled to $H_{0}=70$ km s$^{-1}$ Mpc$^{-1}$, and the crosses in the right panel correspond to {\it Chandra} data \citep{ettori02}. It is remarkable that a cooling catastrophe develops in both systems within about $200$ - $300$ Myr.}
 \label{plot1}
 \end{figure*} 
 
   \begin{figure}
   \centering
\plotone {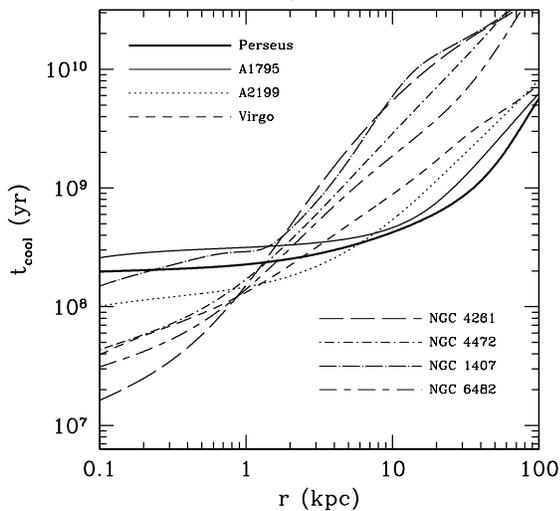} 
\caption{The initial gas cooling time in our simulated systems. For massive clusters (Perseus, Abell 1795 and Abell 2199), the central cooling catastrophe usually happens at about the central gas cooling time. For the other five smaller systems (including Virgo), the gas stays in the hot mode without developing the cooling catastrophe for at least 1 Gyr, much longer than the central gas cooling time, which is often shorter than $0.1$ Gyr.}
 \label{plot2}
 \end{figure} 
 
  \begin{figure*}
   \centering
\plottwo {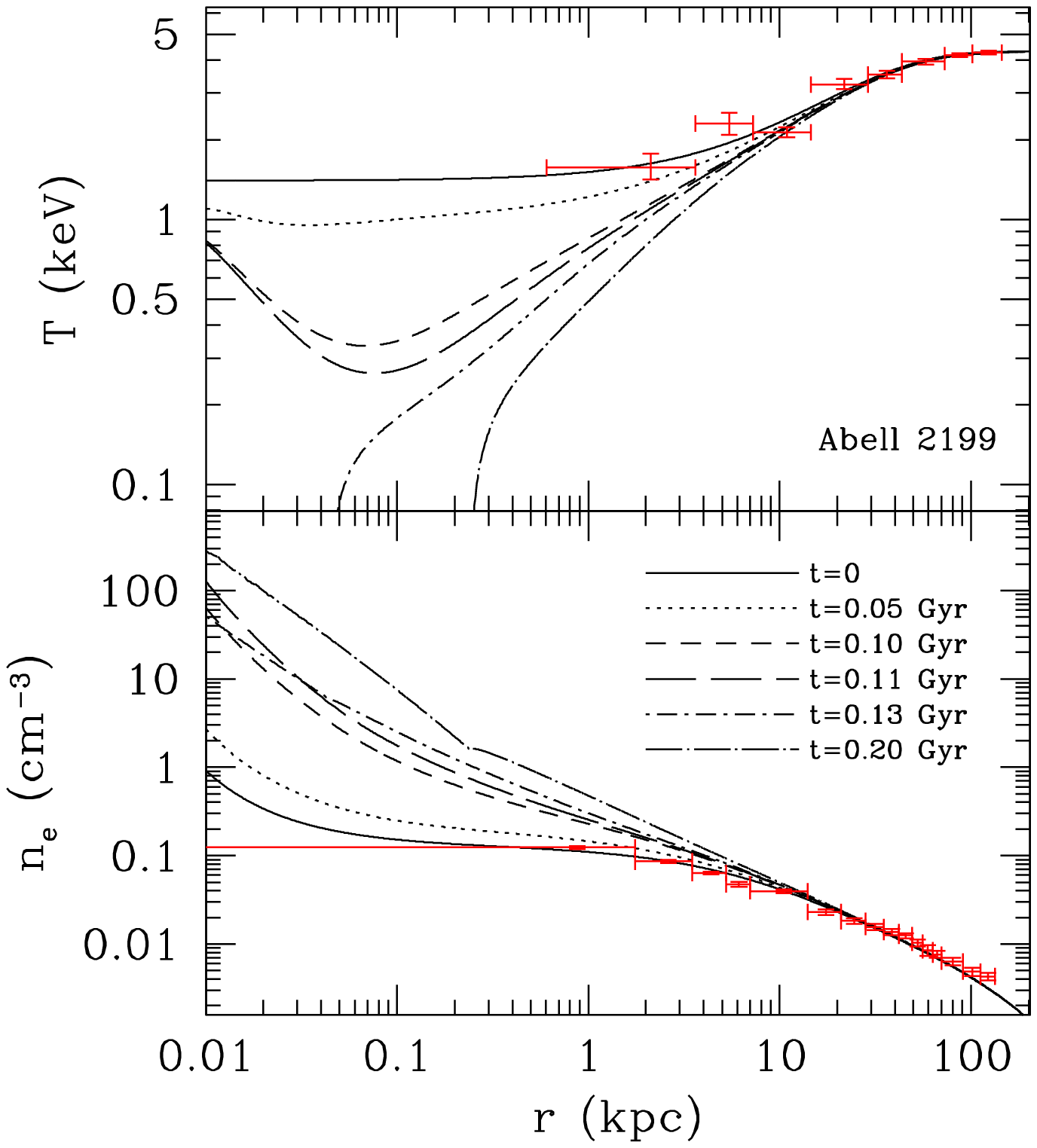}{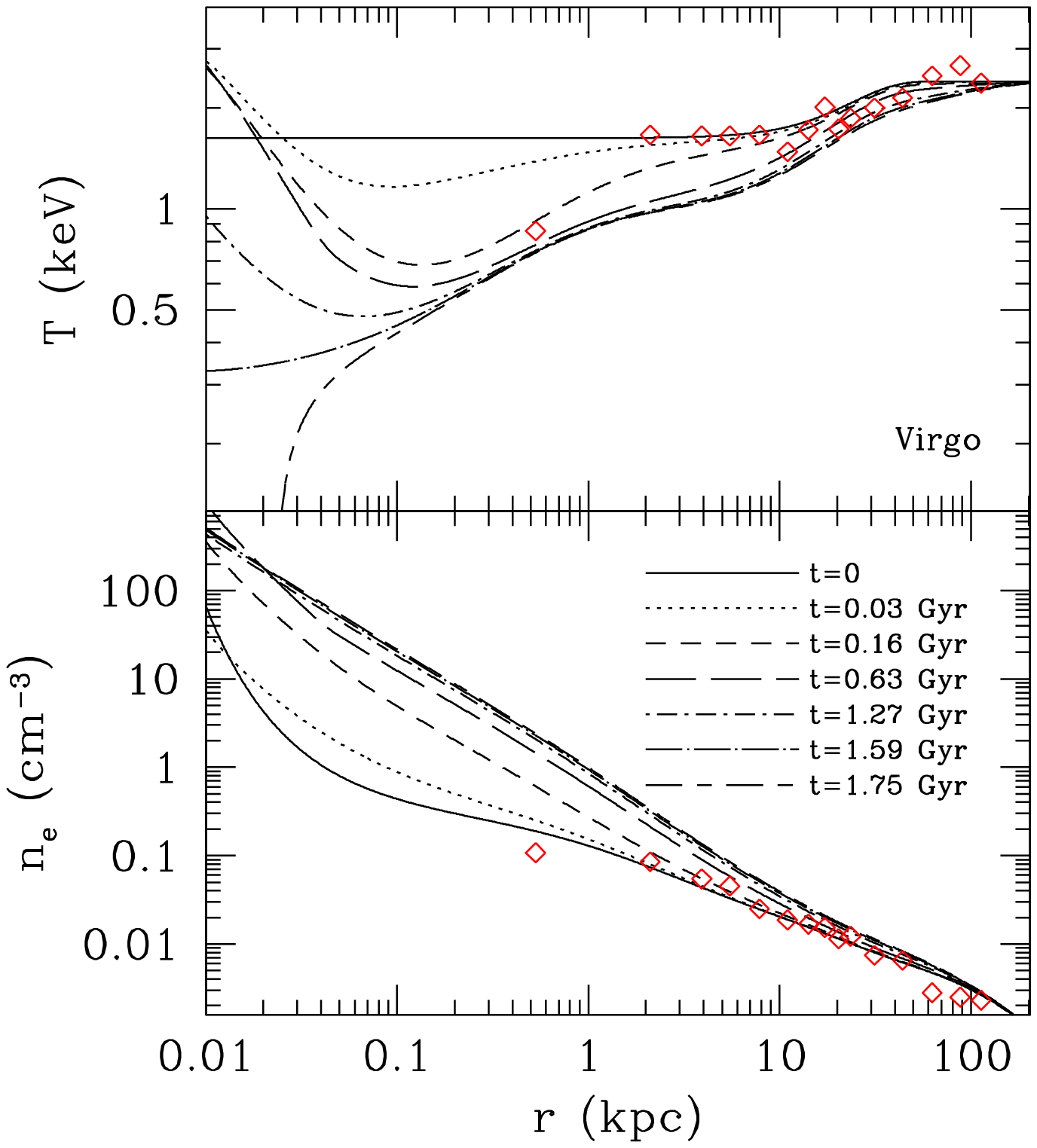} 
\caption{The temporal evolution of temperature and electron number density profiles for Abell 2199 (left) and the Virgo cluster (right). the crosses in the left panel correspond to {\it Chandra} data \citep{johnstone02}, and the diamonds in the right panel are observational data adopted from \citet{ghizzardi04}. In Abell 2199, the cooling catastrophe happens at $t\sim 0.1$ Gyr, while in Virgo, the cooling catastrophe happens at a much later time $t\sim1.7$ Gyr. It is likely that in Virgo, AGN feedback is triggered during the hot-mode accretion, and averts the development of the cooling catastrophe.}
 \label{plot3}
 \end{figure*}

 \section{The Bimodal Transitions from Cooling Flows to Accretion Flows}
\label{section3}

As discussed in detail in the previous section, we assume that the hot gas in each system is initially in hydrostatic equilibrium. The initial temperature and density profiles are chosen to mimic the observational profiles (in regions resolved by current X-ray observations). Without cooling, the hot gas remains in its initial profiles, as confirmed in our simulations. When radiative cooling is turned on at time $t=0$, the gas entropy gradually drops and the gas density increases, resulting in an inflow of hot gas toward the center. Our simulations allow us to study the development of the cooling flow in each real system, and in particular, the transition of the cooling flow to the accretion flow near the central SMBH. 
 
\subsection{The Flow Transition in Galaxy Clusters}

We first study the flow transition in galaxy clusters. The development of cooling flows in massive clusters is very similar, and here we present the results of four typical clusters. In Figure \ref{plot1}, we show the time evolution of the gas temperature and electron number density profiles for two typical cool core clusters: Perseus and Abell 1795. The evolution of these two clusters are very similar: the hot gas initially in hydrostatic equilibrium is perturbed by radiative cooling, which leads to a gradual decrease in the gas entropy and increase in the gas density (in particular at $r\lesssim 10$ kpc). As the gas temperature in these regions drops, a negative temperature gradient develops within about a few tens parsec. The negative temperature gradient is more prominent when the mass of the central SMBH is more massive. This is a signature of the formation of the Bondi-like accretion \citep{bondi52} onto the central SMBH. However, such a temperature profile can not sustain, and the gas temperature in the inner regions continues to decrease due to cooling. At $t\equiv t_{\rm cc} \sim 0.19$ Gyr, a cooling catastrophe happens in the central regions of Perseus, as clearly seen in the dramatic drop of the gas temperature there in the top-left panel of Figure \ref{plot1}. The cooling catastrophe first develops at $r\sim$ several tens pc, and then slowly propagates to larger radii. In Abell 1795, the cooling catastrophe happens at $t_{\rm cc} \sim 0.27$ Gyr. In both systems, the cooling catastrophe happens roughly at the initial gas cooling time at the cluster's central regions, as shown in Figure \ref{plot2}.

This result is consistent with the recent study by \citet{li12}, who investigated the cooling flow to accretion flow transition in the Perseus cluster using three-dimensional simulations. Although they adopted slightly different profiles for both the stellar distribution of the central galaxy and the initial gas density, a cooling catastrophe also develops in their simulations within the central regions of Perseus at $t\sim 300$ Myr. 

The left panel of Figure \ref{plot3} shows the temporal evolution of the gas temperature and electron number density profiles for another massive cluster Abell 2199. The evolution of the cooling flow in this cluster is very similar to that in Perseus and Abell 1795. The cooling catastrophe in Abell 2199 happens at $t\sim 0.1$ Gyr, roughly the initial central gas cooling time. Our simulations of the above three massive clusters suggest that, as the cooling flow is accreted by the central SMBH, the development of a central cooling catastrophe at about the central gas cooling time (typically few hundred Myr) is very robust in massive galaxy clusters, implying that the hot gas cools unimpeded, leading to a cold-mode accretion onto SMBHs. The development of a central cooling catastrophe or central accumulation of cold gas has also been frequently found in previous simulations of cooling flows (e.g., \citealt{brighenti02}; \citealt{brighenti03}; \citealt{cattaneo07}; \citealt{ettori08}; \citealt{gaspari11a}; \citealt{gaspari11b}; \citealt{gaspari12}).

We also studied the flow transition in a less massive cluster, the nearby Virgo cluster, which has gas temperatures of about $2$ - $2.5$ keV in outer regions, much less than those in Perseus ($\sim 6$ keV), Abell 1795 ($\sim 6$ keV) and Abell 2199 ($\sim 4$ keV). The right panel of Figure \ref{plot3} shows the temporal evolution of the gas temperature and electron number density profiles in our Virgo simulation. Compared to the other two clusters, the Virgo cluster develops a larger central region with larger negative temperature gradients during the early times, probably due to stronger gravitational heating by its more massive central SMBH (see Table 1). It is remarkable that the hot gas in Virgo remains in this hot-mode Bondi-like accretion much longer ($> 1$ Gyr), while the initial central gas cooling time is even shorter than 0.1 Gyr. Although the cooling catastrophe still happens in this simulation at $t\sim 1.7$ Gyr, galaxy clusters (including Virgo) are expected to undergo AGN feedback events within much shorter durations (e.g., \citealt{reynolds97}). It is usually thought that AGN feedback events significantly heat the gas, suppressing or even shutting off cooling flows. However, it is also important to note that the development of cooling flows depends on the gas metallicity and a higher metallicity may cause Virgo to undergo the central cooling catastrophe much earlier (see Section 3.3).

  \begin{figure*}
   \centering
\plottwo {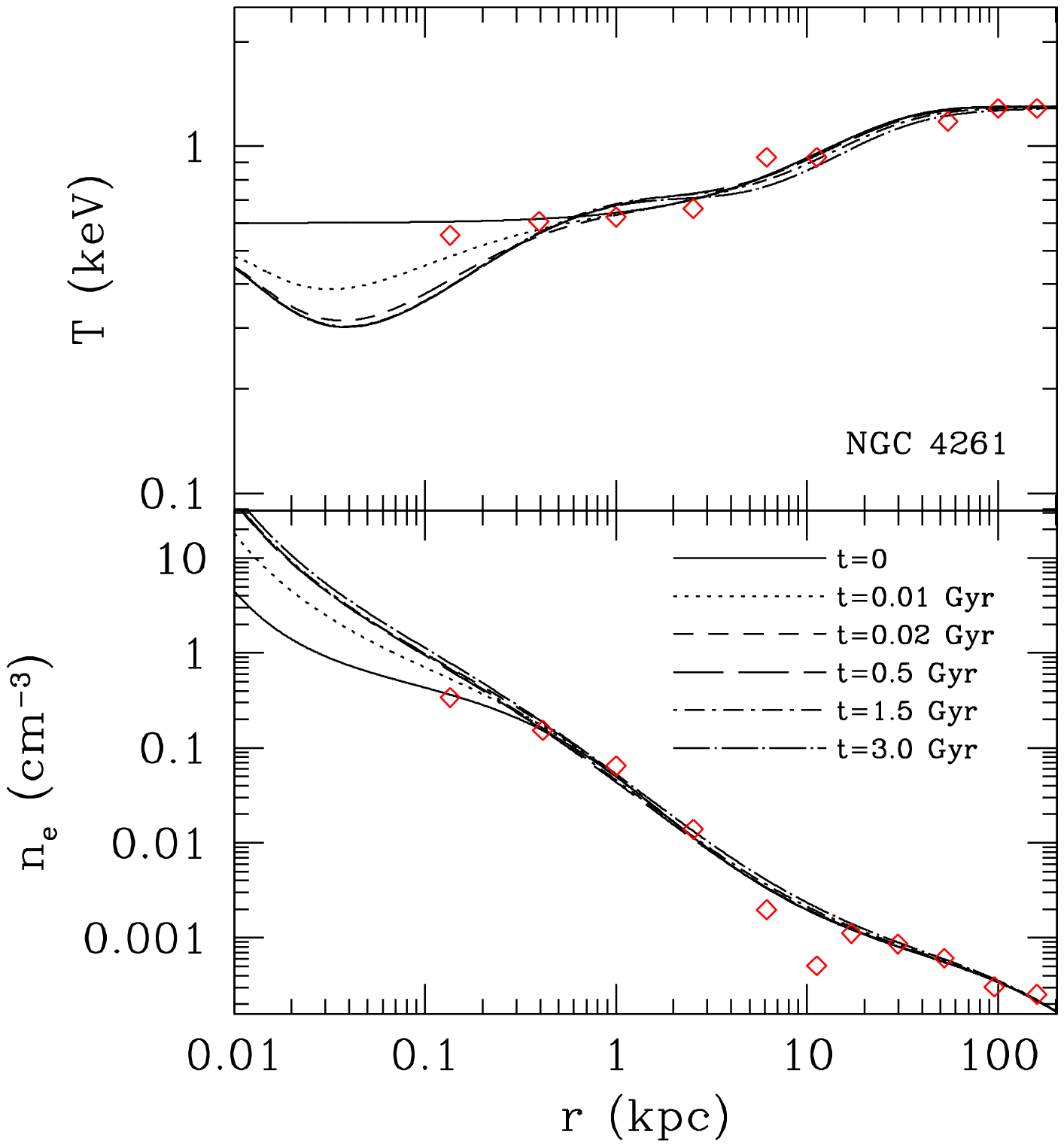} {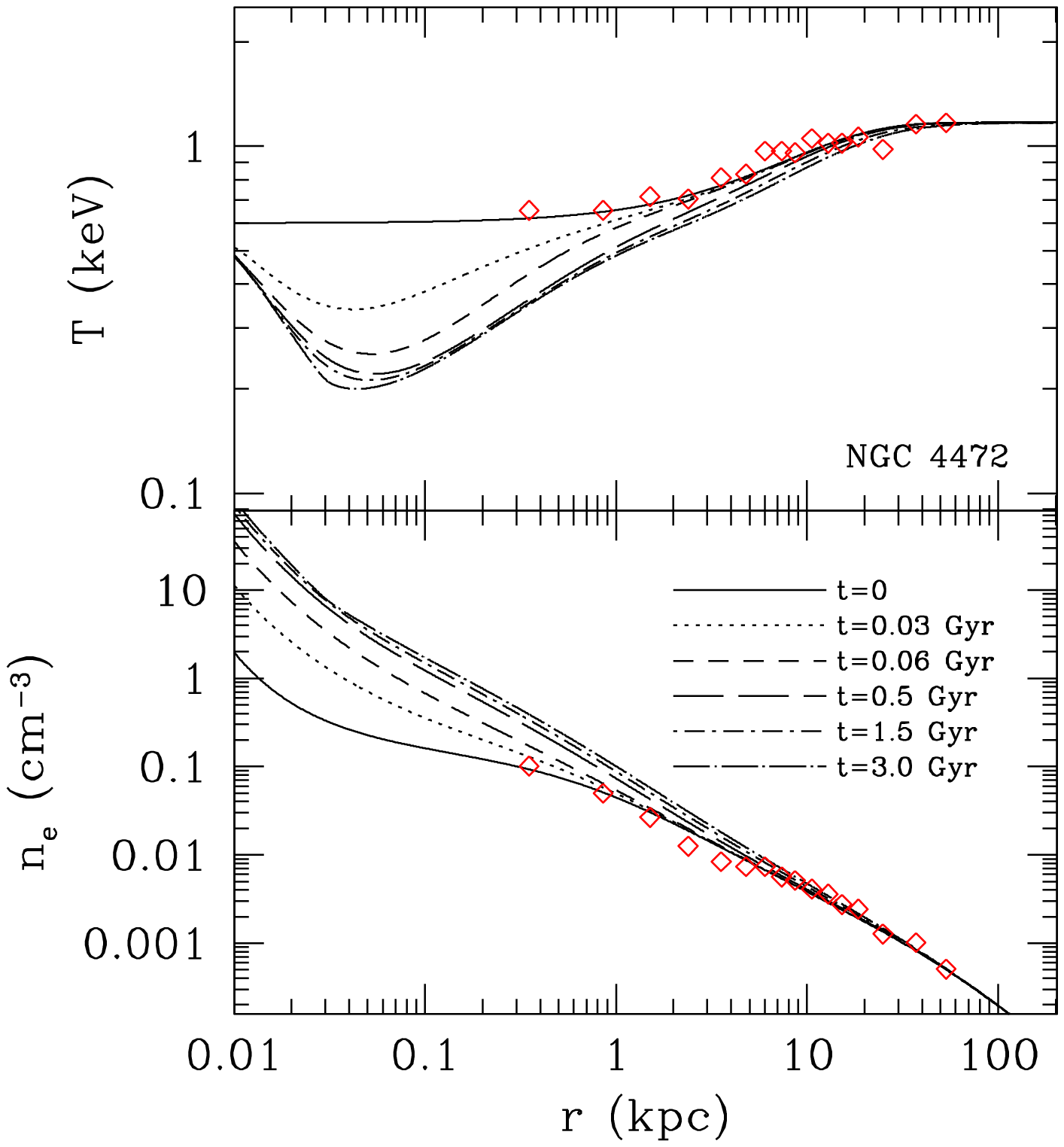}
\caption{The temporal evolution of temperature and electron number density profiles for NGC 4261 (left) and NGC 4472 (right). The diamonds correspond to {\it Chandra} data \citep{humphrey06}. In the top-left panel, the gas temperature remains almost the same from $t\sim 0.02$ Gyr to $3$ Gyr. In both systems, the cooling catastrophe does not happen within $3$ Gyr, indicating a hot-mode accretion for the central SMBH.}
 \label{plot4}
 \end{figure*} 

 \begin{figure*}
   \centering
\plottwo {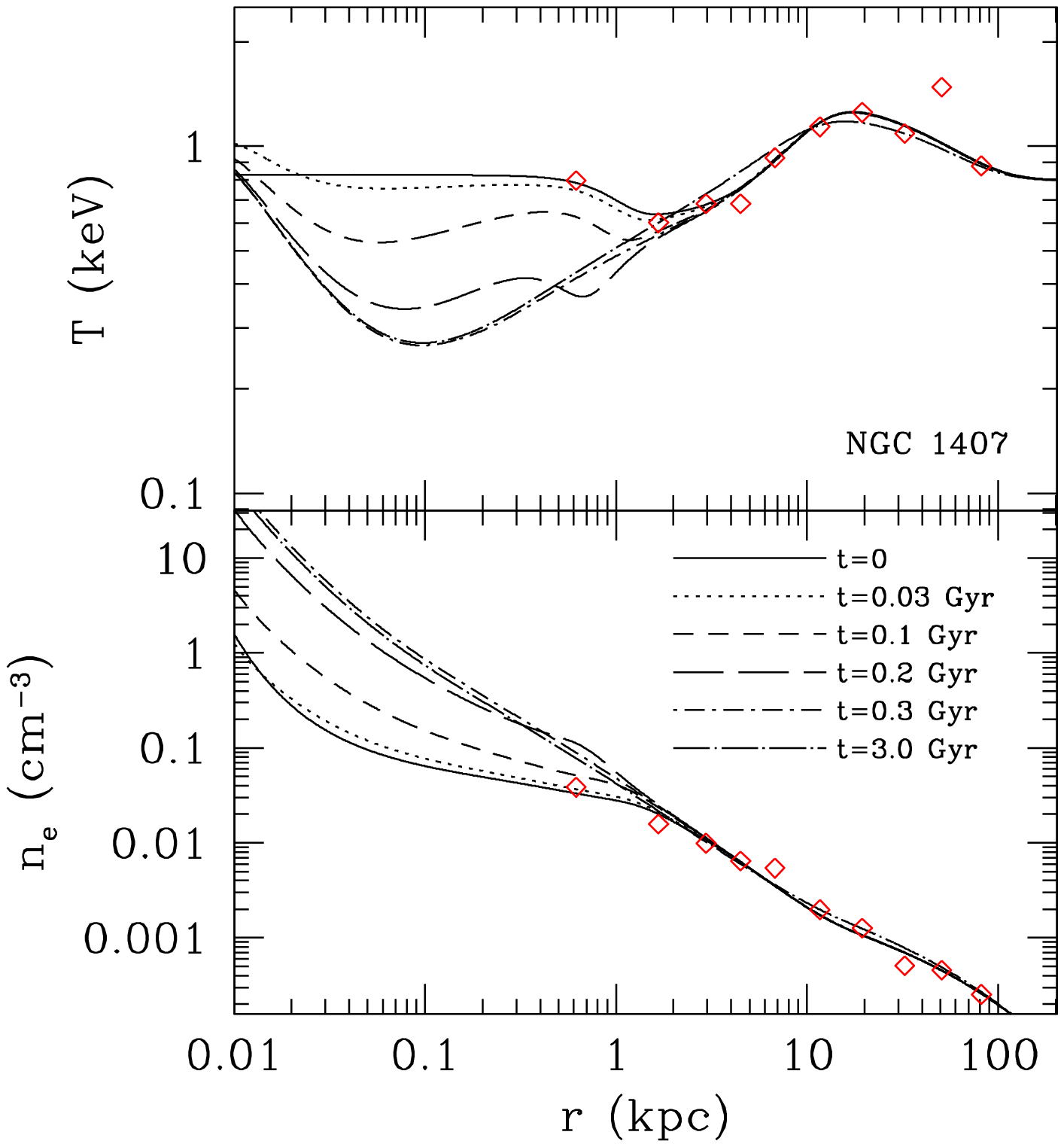} {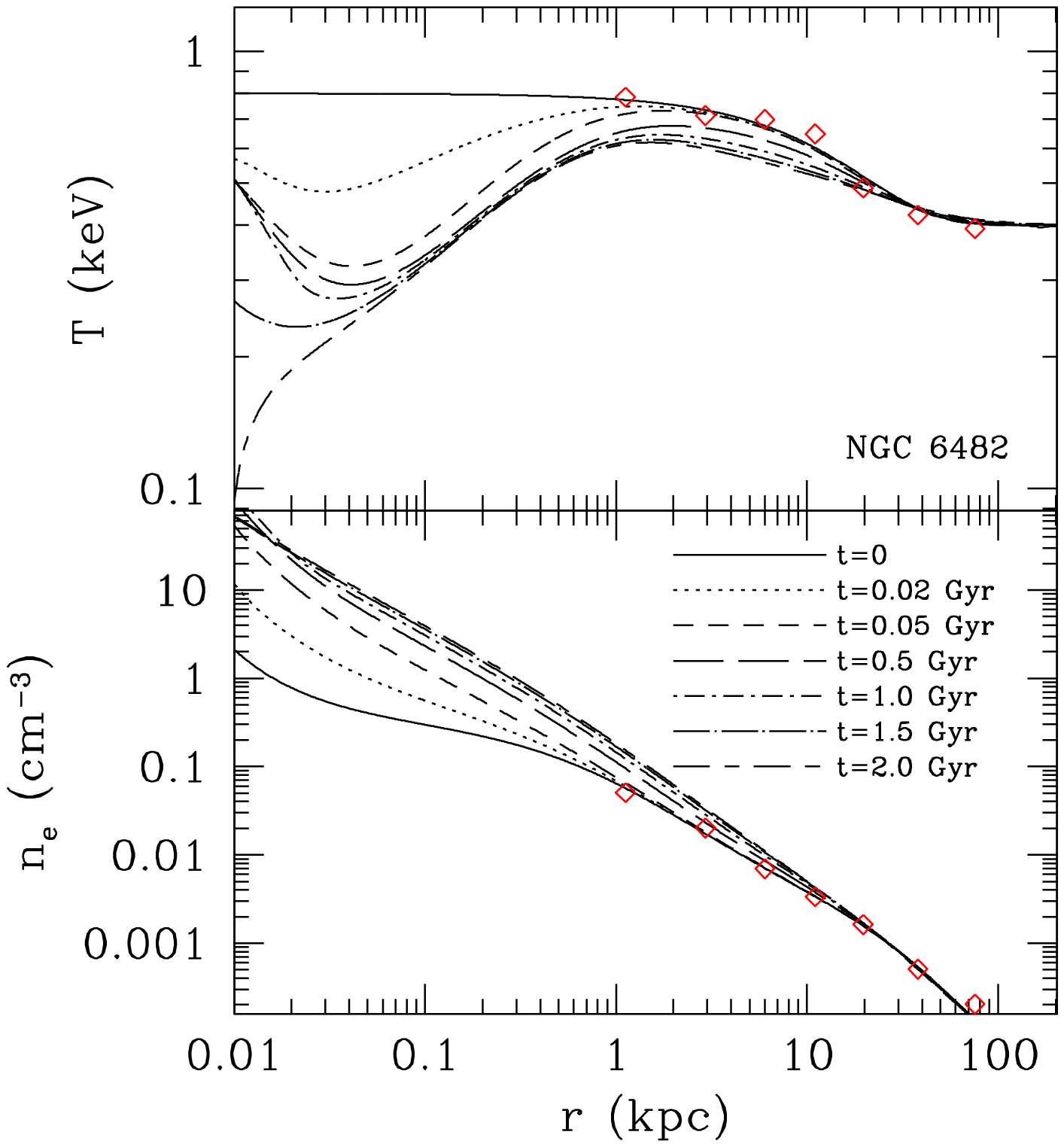}
\caption{The temporal evolution of temperature and electron number density profiles for NGC 1407 (left) and NGC 6482 (right). The diamonds correspond to {\it Chandra} data \citep{humphrey06}. In both groups, the hot gas does not cool off within $3$ Gyr (NGC 1407) or $2$ Gyr (NGC 6482), suggesting a hot-mode accretion for the central SMBH.}
 \label{plot5}
 \end{figure*} 
 
\subsection{The Flow Transition in Elliptical Galaxies and Galaxy Groups}

In smaller systems such as elliptical galaxies and galaxy groups, the cooling flow to accretion flow transition may be quite different compared to massive galaxy clusters. Figure \ref{plot4} shows the temporal evolution of temperature and electron number density profiles for the group NGC 4261 and the elliptical galaxy NGC 4472. Their temperature evolution at early times are similar to galaxy clusters, as described by a gradual decrease in temperature due to cooling and the formation of a cuspy temperature profile with negative temperature gradients in the central regions due to strong compressional heating caused by the SMBH's gravity. However, in galaxy clusters, such a central cuspy temperature profile quickly evolves into a strong cooling catastrophe after about the central cooling time (Fig. \ref{plot1}), while in these two smaller systems, the negative temperature gradient remains in the central regions for a very long time ($>3$ Gyr). For example, in NGC 4261 shown in the left panels of Figure \ref{plot4}, the hot gas reaches a quasi-steady state at $t\sim 0.02$ Gyr (the short-dashed line), and remains in this state at least until the end of our simulation ($t=3$ Gyr; the dot-long-dashed line). This is remarkable as the central gas cooling time in NGC 4261 is very short (less than 100 Myr). This quasi-steady state is featured by a hot Bondi-like accretion in the central regions (within about $50$ pc) with negative temperature gradients. The similar quasi-steady state is also seen in the simulation of NGC 4472 (the right panels of Fig. \ref{plot4}). Within such a long duration, AGN feedback is very likely triggered during the hot-mode accretion. This suggests that the cooling flow in these two systems transitions into a hot-mode accretion onto the central SMBH.

We experimented with two additional galaxy groups (NGC 1407 and NGC 6482), and found similar results. Figure \ref{plot5} shows the time evolution of temperature and electron number density profiles for these two groups. In NGC 1407, the gas reaches a hot-mode quasi-steady state at $t\sim 0.3$ Gyr and stays in this state until at least $t=3$ Gyr. Compared to other groups, NGC 1407 develops a larger central region with negative temperature gradients ($r\lesssim 100$ pc), due to its more massive SMBH. NGC 6482 is a fossil group with an observational negative temperature gradient from about 1 kpc to about 100 kpc. In the simulation for this group, the hot-mode accretion in the central regions also sustains for a very long time ($>1$ Gyr) and the cooling catastrophe only happens at about $t\equiv t_{\rm cc}=2$ Gyr. During such a long timescale, it is very likely that AGN feedback events will be triggered during the hot-mode accretion, preventing the otherwise onset of the cooling catastrophe at $t_{\rm cc}=2$ Gyr (similar to the Virgo cluster).

\subsection{Dependence on the Gas Metallicity}

   \begin{figure}
   \centering
\plotone {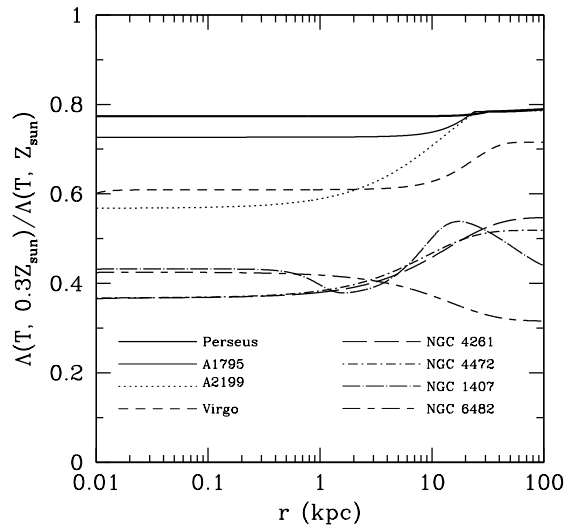} 
\caption{The impact of metallicity on the gas cooling rate. Plotted are the ratios of the initial gas cooling rates at $Z=0.3Z_{\sun}$ to those at $Z=Z_{\sun}$ for all our simulated systems. The impact is relatively small for galaxy clusters, but becomes quite significant for galaxy groups and ellipticals (a typical increase of the cooling rate by a factor of $2$ - $3$ when increasing the metallicity from $0.3Z_{\sun}$ to $Z_{\sun}$.)}
 \label{plot6}
 \end{figure} 
 
 \begin{figure*}
   \centering
\plottwo {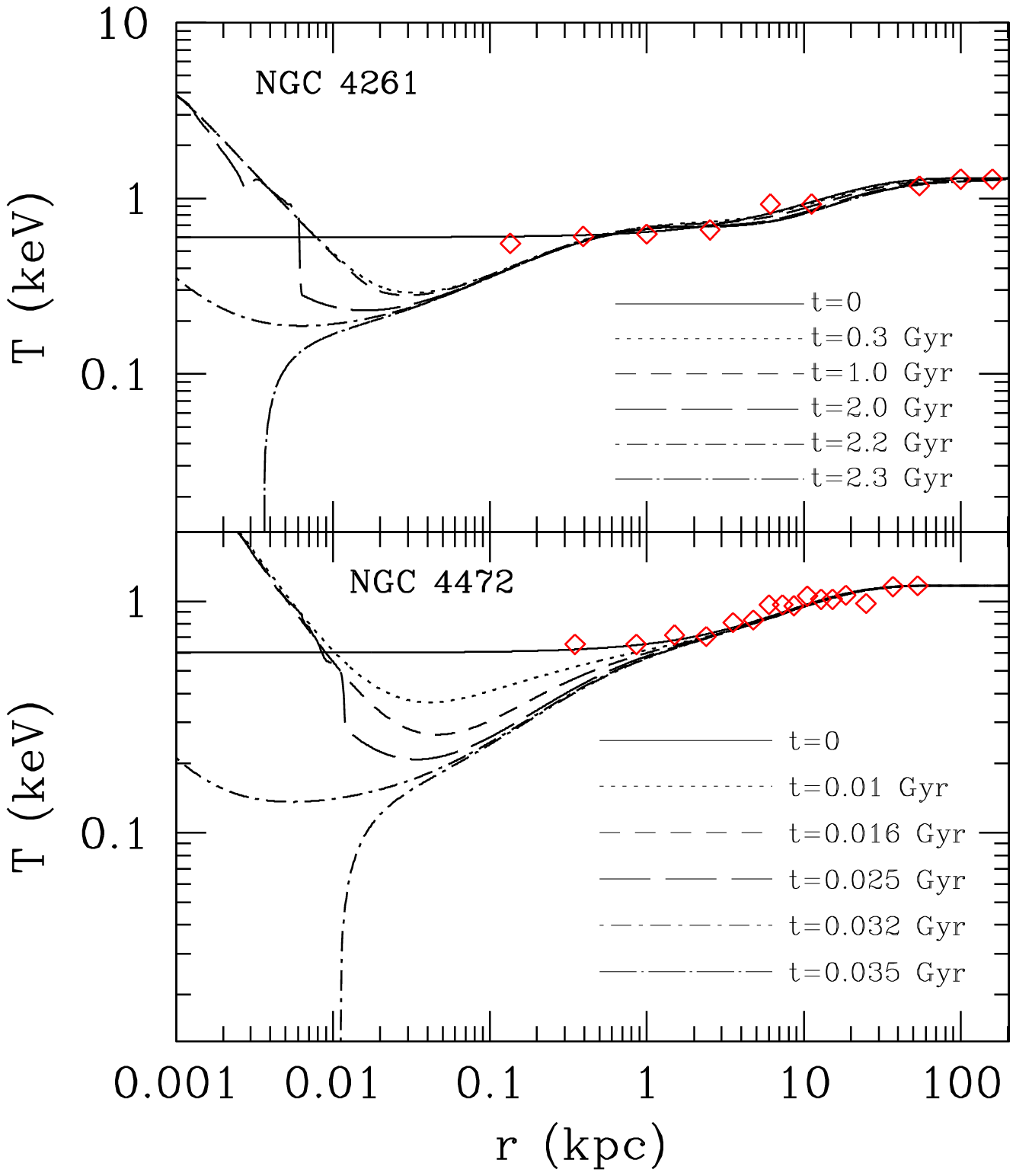} {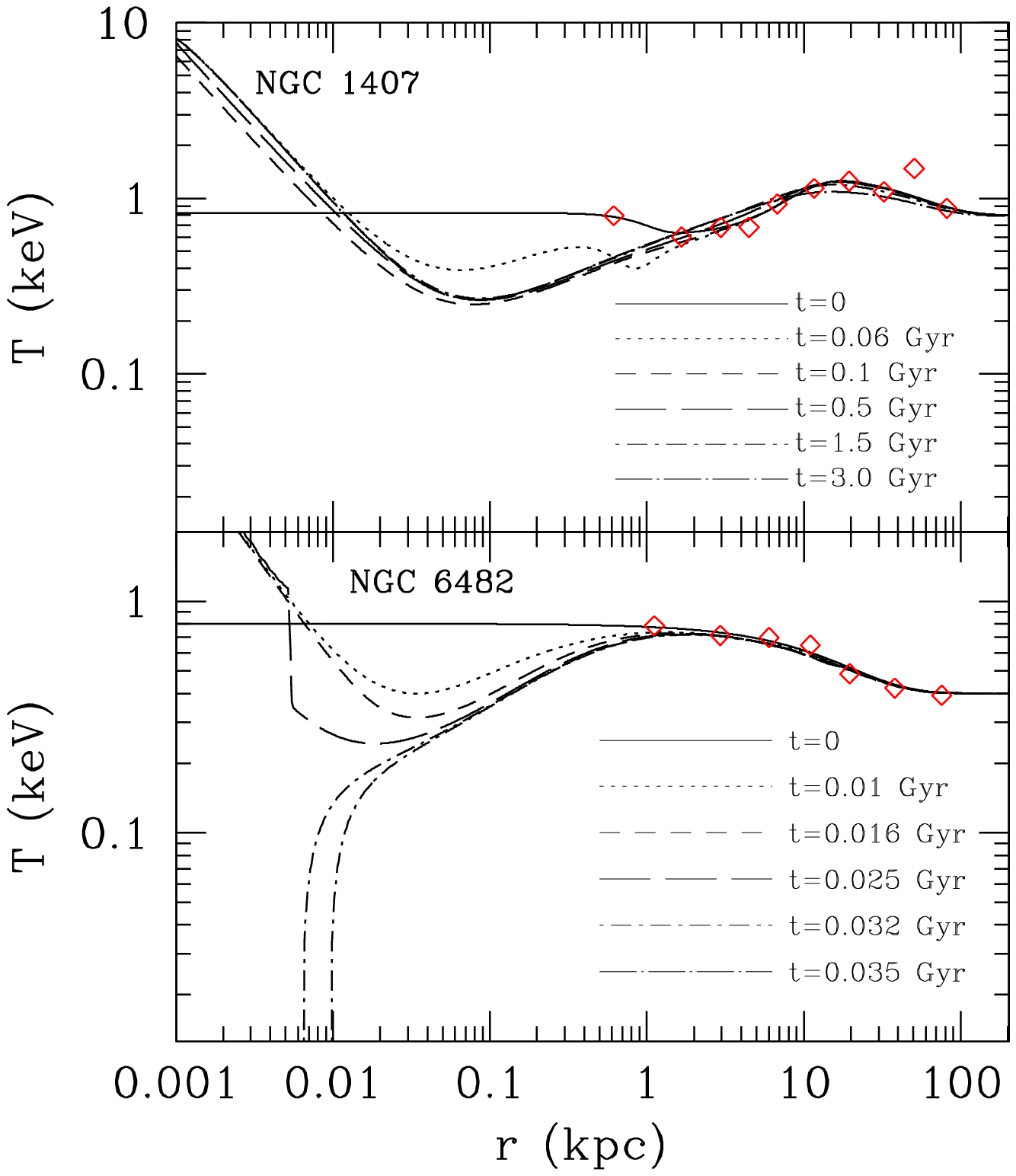}
\caption{The temporal evolution of the gas temperature profiles in a new set of simulations with the gas metallicity $Z=Z_{\sun}$ and covering a radial range from $1$ pc to $200$ kpc  for the four relatively small systems in our sample. The higher metallicity increases the gas cooling rate, resulting into the fast development of the central cooling catastrophe in NGC 4472 and NGC 6482 at $t\sim 30$ Myr, while the accretion in NGC 4261 and NGC 1407 remains in the hot mode for at least 2 Gyr. The diamonds correspond to {\it Chandra} data \citep{humphrey06}.}
 \label{plot7}
 \end{figure*} 
 
In this subsection, we study how our results depend on the gas metallicity, which affects the cooling rate $\Lambda(T, Z)$ and potentially the development of cooling flows, especially in low-temperature systems such as galaxy groups and galaxies. To this end, we performed a new set of simulations with the \citet{sd93} cooling function at the gas metallicity $Z=Z_{\sun}$ for the four relatively small systems in our sample (NGC 4261, 4472, 1407 and 6482). These simulations are performed with a large spatial range, covering $r_{\rm min}=1$ pc to $r_{\rm max}=200$ kpc. The dependences of our results on the choices of the inner and outer boundaries are explored in Appendix A.

In Figure \ref{plot6}, we show the ratios of the initial gas cooling rates at $Z=0.3Z_{\sun}$ to those at $Z=Z_{\sun}$ for all our simulated systems. The metallicity's impact on the cooling rate is relatively small in our four galaxy clusters (covering a temperature range from about 2 - 7 keV) , increasing the cooling rate by a factor of about 1.2 - 1.6. But for our four smaller systems at $T \sim 1$ keV, the factor is about 2 - 3, consistent with the fact that metal cooling is more important in colder systems \citep{sd93}.

Figure \ref{plot7} shows the temporal evolution of the gas temperature profiles in this new set of simulations with $Z=Z_{\sun}$. In our previous simulations of these four small systems at $Z=0.3Z_{\sun}$, all of them develop a hot-mode accretion in the central regions sustaining for at least 2 Gyr. When cooling is enhanced with $Z=Z_{\sun}$ in the new simulations, two systems, namely NGC 4472 and NGC 6482, encounter a central cooling catastrophe quickly at $t\sim 30$ Myr, resulting in a cold-mode accretion onto the SMBHs. We have also run the simulation for the Virgo cluster with the new cooling function at $Z=Z_{\sun}$, showing that the central cooling catastrophe happens at $t_{\rm cc}\sim 0.22$ Gyr, much earlier than $t_{\rm cc}\sim 1.7$ Gyr in the original simulation with $Z=0.3Z_{\sun}$.

However, for the other two systems, NGC 4261 and NGC 1407, the hot gas in the central accretion region remains in the hot mode with negative temperature gradients for at least 2 Gyr. This is remarkable as the central gas cooling times in these two systems are much less than 0.1 Gyr.

In summary, higher metallicity increases the gas cooling rate, more notably in lower-temperature systems, e.g., galaxies and galaxy groups, leading to the central cold-mode accretion of cooling flows in some systems, which would otherwise host the hot-mode accretion if at a lower metallicity. However, even at the solar metallicity, some systems (e.g., NGC 4261 and NGC 1407) still host the hot-mode SMBH accretion of cooling flows. In real cool core systems, the metallicity distribution is usually not uniform, but instead drops with radius (e.g., \citealt{degrandi01}; \citealt{rp07}). In galaxy groups, the central gas metallicity is often near solar (higher and lower metallicities also exist in some systems). But it usually drops to $Z\sim 0.1Z_{\sun}$ within $r\sim r_{500}$, where $r_{500}$ is the radius enclosing a mean density of 500 times the critical density \citep{rp07}. Thus the metallicity impact in real systems is more complicated, but our two sets of simulations with $Z=0.3Z_{\sun}$ and $Z=Z_{\sun}$ indicate convincingly that some galaxy groups and elliptical galaxies host the hot-mode SMBH accretion of cooling flows when other more subtle physics, e.g., AGN heating, turbulence, stellar mass losses, are not important.

\subsection{The Dichotomy in the Flow Transition}

  \begin{figure}
   \centering
\plotone {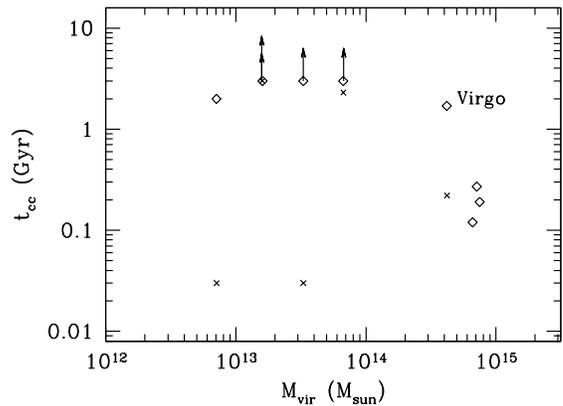} 
\caption{The time when the cooling catastrophe happens (since the beginning of each simulation) $t_{\rm cc}$ versus the virial mass of the dark matter halo ($M_{\rm vir}$) in our simulations of the eight systems. Diamonds correspond to our standard simulations with $Z=0.3Z_{\sun}$, while the five small systems are also studied by a new set of simulations with $Z=Z_{\sun}$ (crosses). The arrows attached to diamonds or crosses indicate that no cooling catastrophe develops during the whole simulation, which stops at $t=3$ Gyr (i.e., $t_{\rm cc}>3$ Gyr). For NGC 1407, $t_{\rm cc}>3$ Gyr in both simulations.}
 \label{plot8}
 \end{figure} 
 
In this subsection, we summarize our main findings in the previous three subsections. The hot gas in our calculations is initially in hydrostatic equilibrium, and thus does not evolve without cooling. When radiative cooling is turned on, the gas gradually cools, especially in the inner regions where the cooling time is relatively short. In the central regions, a cuspy temperature profile with negative temperature gradients develops due to strong compressional heating caused by the SMBH's gravity. The spatial size of such central regions depends on the mass of the central SMBH, varying in our sample from a few tens to about 100 pc (e.g., Virgo, NGC 1407), and may be even larger if the central SMBH is more massive.

More interestingly, our simulations show that the further development of cooling flows, in particular the central accretion region with negative temperature gradients, is bimodal. In galaxy clusters and some smaller systems, the central hot-mode accretion does not sustain, but instead evolves quickly into a cooling catastrophe at about the central gas cooling time, which is about $100$ - $300$ Myr for galaxy clusters and much smaller for galaxy groups and elliptical galaxies, depending on the gas metallicity. In contrast, in some elliptical galaxies and galaxy groups, the hot-mode accretion becomes quasi-steady, sustaining for at least 2 Gyr or even permanently.

This dichotomy is clearly summarized in  Figure \ref{plot8}, which shows $t_{\rm cc}$, the time when the cooling catastrophe happens after the beginning of each simulation, versus the virial mass of the dark matter halo for our eight systems in two sets of simulations with different gas metallicities. Assuming a constant gas metallicity of $Z=0.3Z_{\sun}$ (diamonds in Figure \ref{plot8}), the three most massive systems develop the central cooling catastrophe at about their central gas cooling times ($t_{\rm cc} \sim 100$ - $300$ Myr), while the other five smaller systems host the central hot-mode accretion for $t\sim t_{\rm cc} \gtrsim 2$ Gyr, though their central gas cooling times are typically about 100 Myr or shorter. As the gas metallicity is increased to $Z=Z_{\sun}$ (crosses in Figure \ref{plot8}), three of the five smaller systems develop the central cooling catastrophe quickly, but the other two, namely NGC 4261 and NGC 1407, still host the central hot-mode accretion, which is quasi-steady, lasting for about 2 Gyr or longer. 
Higher metallicity increases the incidence of cold-mode accretion, but as discussed in Section 3.3, our calculations indicate that the hot-mode accretion of cooling flows does operate in some elliptical galaxies and galaxy groups quasi-steadily unless AGN feedback or other physical processes significantly disturb cooling flows, changing our results. 
  
 \section{The Origin of the Dichotomy}
\label{section4}

  \begin{figure*}
   \centering
\plottwo {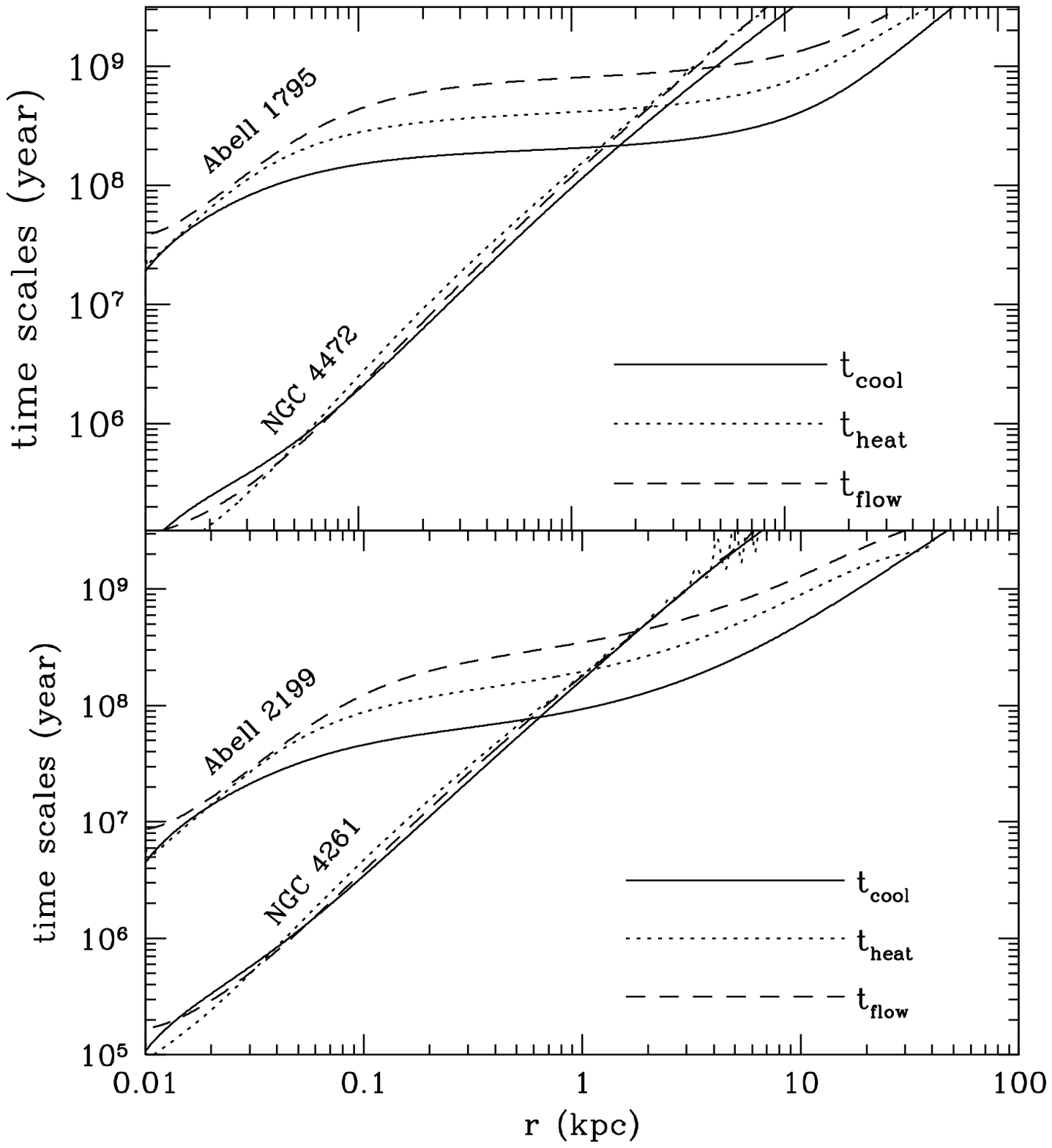} {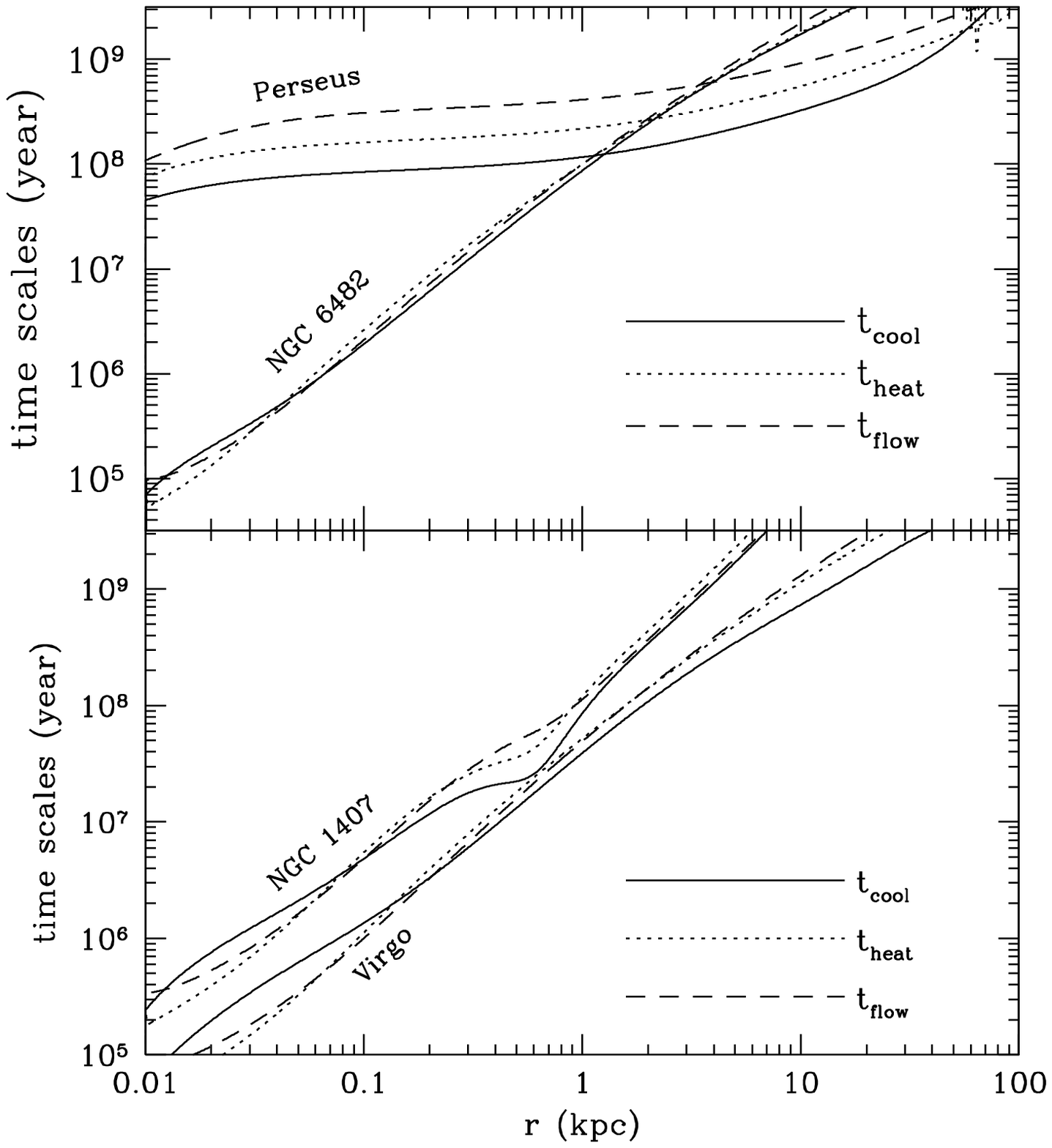} 
\caption{The cooling time ($t_{\rm cool}$), compressional heating time ($t_{\rm heat}$), and inflow time ($t_{\rm flow} \equiv r/|v|$) in our standard simulations of the eight systems. These timescales evolve with time during the simulations. For the three massive clusters, they are plotted at a representative time before reaching the cooling catastrophe: $t=0.1$ Gyr  for Abell 1795 and Perseus, $t=0.05$ Gyr for Abell 2199, while for the other five smaller systems, they are plotted at $t=0.2$ Gyr, a typical time when the system has already reached the quasi-steady state.}
 \label{plot9}
 \end{figure*} 

In this section, we discuss the possible origin of the dichotomy in the transition of cooling flows to accretion flows discovered in the previous section. In Section 4.1, we study various important physical processes that operate during the development of cooling flows, focusing on radiative cooling, compressional heating, and gas inflow. We investigate how SMBH accretion flows fed by cooling inflows are affected by the initial gas density distribution and the gas temperature in Section 4.2, and the importance of the gravity from the central galaxy in Section 4.3.

\subsection{Cooling, Compressional Heating and Gas Inflow}

To understand the origin of the dichotomy, here we qualitatively explore the physical processes operating during the development of cooling flows. In addition to radiative cooling, the hot gas is also subject to adiabatic compression, represented by the term $-P\nabla \cdot {\bf v}$ in equation \ref{hydro3}. Adiabatic compression heats the hot gas through the $pdV$ work, and this energy comes from the release of the gravitational energy of the infalling gas. For an arbitrary small gas parcel in a cooling flow, the evolution of its internal energy is determined by the interplay of radiative cooling and compressional heating (equation \ref{hydro3}). However, even when cooling dominates over heating, the {\it energy density} of the gas parcel usually still increases with time, as its spatial volume decreases due to compression. The impact of compressional heating may be better described in the evolution of the gas temperature:
\begin{eqnarray}
\frac{e}{T}\frac{dT}{dt}=-P\nabla \cdot {\bf v}-n_{\text{e}}n_{\text{i}}\Lambda(T, Z) 
   \rm{ ,}\label{tevolution}
   \end{eqnarray}
which can be derived directly by combining equations (\ref{hydro1}) and (\ref{hydro3}). Equation (\ref{tevolution}) states that, if  radiative cooling dominates over compressional heating, the temperature of the gas parcel drops with time as it flows toward the center. On the other hand, if compressional heating rate is higher than cooling rate, the temperature of the gas parcel increases with time as it flows inward.

At small radii, the compressional heating rate ($\Gamma_{\rm comp}= -P\nabla \cdot {\bf v} \sim 2Pv/r$, where $v=|{\bf v}|=-v_{\rm r}$) is very large, due to the SMBH's strong gravity and the small values of $r$. As cooling flows develop in our simulations, the compressional heating rate usually surpasses cooling rate very quickly at small enough radii (typically $r\lesssim 10$ - $100$ pc, depending on the SMBH's mass), where the gas temperature may rise toward the center, leading to a central negative temperature gradient as shown in Figures \ref{plot1}, \ref{plot3}, \ref{plot4}, \ref{plot5}, and \ref{plot7} (but a central cooling catastrophe develops at a later time in some systems). 

In contrast, at larger radii, cooling rate is usually larger than the compressional heating rate in our simulations. Thus for a gas parcel at large radii but still within the effective cooling region (where the cooling time is shorter than a few Gyr), its temperature drops with time as it flows inward. Once the gas parcel reaches the central region where the heating rate surpasses the cooling rate, its temperature starts to increase with time, resulting in a hot-mode accretion for the central SMBH. On the other hand, if the gas parcel cools off (i.e. cool to zero temperature) before reaching the central region, we get a cold-mode accretion for the SMBH. Thus the fate of cooling flows is determined by three important physical mechanisms: radiative cooling, gas inflow and compressional heating. The former two can also be seen through the time evolution of the gas entropy $S$ (defined in equation \ref{sdefinition}):
\begin{eqnarray}
\frac{en_{\rm e}^{2/3}}{k_{\rm B}T}\frac{dS}{dt}=-n_{\text{e}}n_{\text{i}}\Lambda(T, Z) 
   \rm{ ,}\label{sevolution}
   \end{eqnarray}
which can be derived from equations (\ref{hydro1}), (\ref{hydro3}), and (\ref{sdefinition}), and states that the entropy of a gas parcel always drops as it flows toward the center due to radiative cooling. 
   
It is important to note that these three physical mechanisms do not operate separately. Cooling induces gas inflow and compressional heating, and they are strongly entangled during the development of cooling flows, depending on the gas properties and the gravitational potential well. Their relative importance can be quantitatively seen in Figure \ref{plot9}, which shows the cooling time $t_{\rm cool}$ (defined in equation \ref{tcooldef}), the compressional heating timescale
\begin{eqnarray}
t_{\rm heat}=e/(-P \nabla \cdot {\bf v})  
\end{eqnarray}
and the inflow timescale $t_{\rm flow} \equiv r/|v|$ in our standard simulations (presented in Sections 3.1 and 3.2) of the eight systems at various simulation times. Let us first look at NGC 4472 shown in the top-left panel. The three timescales are plotted at $t=0.2$ Gyr, a representative time when the hot gas in NGC 4472 is already in the quasi-steady state. Within the central several tens pc, the compressional heating time is the shortest timescale (explaining the negative temperature gradient seen in the right-top panel of Figure \ref{plot4}), while outside this central region, the cooling time is slightly shorter than the other two timescales. However, the three timescales are very close to each other across almost the entire cooling regions. The closeness of these three timescales is quickly built up from small to large radii during the development of cooling flows, and is also seen in the standard simulations of NGC 4261, NGC 1407, NGC 6482, and Virgo, explaining why the hot gas in our simulations of these systems can stay in a quasi-steady state for several Gyr or longer. 

In contrast, the relative magnitudes of these three timescales in our simulations of massive clusters (Perseus, Abell 1795 and Abell 2199) are very different. Figure \ref{plot9} also shows these three timescales for these three systems at a typical time before the onset of the cooling catastrophe ($t=0.1$ Gyr for Perseus and Abell 1795, and $t=0.05$ Gyr for Abell 2199). As in groups, the compressional heating time is often the shortest timescale at the very central regions as the cooling flow develops (except for Perseus, whose central SMBH mass is relatively small), but in almost the entire cooling regions beyond this central region, the gas cooling time is {\it significantly} shorter than the heating and inflow timescales. It is thus not surprising that these systems reach a cooling catastrophe quickly (after about the initial gas cooling time). 

One can also see the relative magnitudes of $t_{\rm heat}$, $t_{\rm cool}$, and $t_{\rm flow}$ through
\begin{eqnarray}
\frac{t_{\rm flow}}{t_{\rm cool}}=\frac{r/v}{e/(n_{\text{e}}n_{\text{i}}\Lambda(T,Z))}\propto \frac{r\rho\Lambda(T,Z)}{vT}  {\rm ,}\label{tscale1}
\end{eqnarray}
and
\begin{eqnarray}
\frac{t_{\rm heat}}{t_{\rm cool}}=\frac{n_{\text{e}}n_{\text{i}}\Lambda(T,Z)}{-P \nabla \cdot {\bf v}}
\propto \frac{r\rho\Lambda(T,Z)}{vT}  {\rm ,}    \label{tscale2}
\end{eqnarray}
where we approximated $-\nabla \cdot {\bf v}\propto v/r$. The timescale ratios in the above two equations depend on a combination of parameters, including the gas density, temperature, velocity, and metallicity. Higher metallicity leads to larger cooling rate, increasing the importance of cooling, as explored in detail in Section 3.3. Higher gas densities correspond to shorter cooling times and larger values of $t_{\rm flow}/t_{\rm cool}$ and $t_{\rm heat}/t_{\rm cool}$, leading to the relatively more important role of radiative cooling (see Section 4.2). The evolution of gas velocity is significantly affected by the gravitational acceleration, as the increase of the kinetic energy of a gas parcel in the cooling flow comes from its gravitational energy, part of which is also converted into the internal energy (see Section 4.3). The impact of the gas temperature is more subtle. On one hand, lower temperature leads to larger values of $t_{\rm flow}/t_{\rm cool}$ and $t_{\rm heat}/t_{\rm cool}$, corresponding to the relatively more important role of radiative cooling. On the other hand, lower temperature results in gas densities dropping faster radially in hydrostatic configuration, leading to lower gas densities and the less important role of radiative cooling. The latter impact becomes more significant in massive clusters, leading to the central cooling catastrophe even at a low metallicity $Z=0.3Z_{\sun}$ (see Section 3.1). We will further study the role of gas temperature and density in Section 4.2.

Although the cooling catastrophe happens in central regions of some systems (see Figs. \ref{plot1} and \ref{plot3}), it should not be only viewed as a {\it local} phenomenon. Instead, it is a {\it global} cooling catastrophe, resulted from strong gas cooling over a large radial range with short cooling timescales (up to tens or few hundreds kpc in massive clusters). Cooling flows at kpc and larger radii bring gas inward, while at the very central regions, the gas is gravitationally heated by the SMBH's potential well, limiting the growth of the mass inflow rate. Thus the hot gas gradually accumulates in central regions (at $r\sim$ a few to tens pc), eventually leading to the cooling catastrophe. 

In summary, the development of cooling flows is determined by the interplay between radiative cooling, gas inflow, and compressional heating. The bimodal fate of cooling flows near the central SMBHs found in our simulations (Sec. 3) is due to the dichotomy in the relative importance of radiative cooling. In massive clusters, the cooling time is significantly shorter than the gas inflow time and the compressional heating time across almost the whole cooling regions (except the very central regions), and the hot gas suffers from a central cooling catastrophe after about the central gas cooling time. In contrast in some smaller systems (galaxy groups and elliptical galaxies), the hot gas quickly reaches a quasi-steady state, where the cooling time, compressional heating time and inflow time are very close to each other. Although the central gas cooling time is very short (e.g., $t_{\rm cool} \sim 0.1$ Gyr at 1 kpc, and even shorter at smaller radii; see Fig. \ref{plot2}), the hot gas can stay in this quasi-steady state for several Gyr or longer. The origin of the bimodality in the relative importance of radiative cooling is related to the initial gas density distribution, the gas temperature, and the gravitational potential well. We explore them in the following two subsections.

\subsection{The Gas Density Profile and Temperature}
  \begin{figure}
   \centering
\plotone {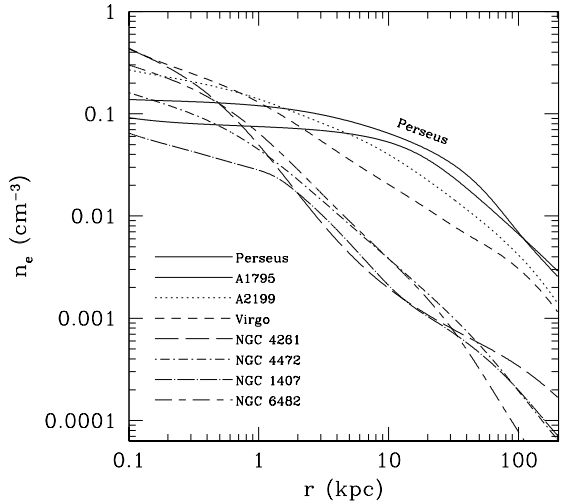} 
\caption{Initial gas density profiles of the eight systems presented in the paper. At $r>1$ kpc, the gas density drops much slower in galaxy clusters compared to smaller systems. Thus gas densities at $r>1$ kpc are much higher in galaxy clusters than in galaxy groups and elliptical galaxies.}
 \label{plot10}
 \end{figure}

The gas cooling rate depends strongly on the gas density. Higher gas densities correspond to larger ratios of $t_{\rm flow}/t_{\rm cool}$ and $t_{\rm heat}/t_{\rm cool}$, leading to a stronger relative importance of radiative cooling. Furthermore, higher gas densities on kpc and larger scales result in larger cooling inflow rates (as in massive clusters), and the gas may accumulate very quickly in the $10 \lesssim r \lesssim 100$ pc region if the growth of the central gas inflow rate is limited by the strong compressional heating due to the SMBH. Thus the gas density distribution is expected to play a significant role in the fate of cooling flows near the SMBHs.
 
Figure \ref{plot10} shows the initial gas density profiles of the eight systems presented in the paper. At $0.1<r<1$ kpc, all the systems have similar gas densities ($\sim 0.1$ cm$^{-3}$). But beyond $1$ kpc, a bimodality clearly shows up: gas densities in galaxy clusters are much higher than those in galaxy groups and elliptical galaxies. This is mainly because that at $r>1$ kpc, the gas density in galaxy groups and elliptical galaxies drops much faster than that in massive clusters. In other words, the scale length of the density profile is much shorter in smaller systems. Assuming that the gas is in hydrostatic equilibrium and isothermal, the scale length of the gas density profile is
\begin{eqnarray}
l_{\rho}\equiv \frac{\rho}{d\rho/dr} \propto \frac{T}{g} \text{ .} \label{lrho}
\end{eqnarray}
Thus the density scale length scales with the gas temperature, and scales inversely with the gravitational acceleration. Massive clusters have higher gas temperatures and thus longer density scale lengths. This partially explains why the gas density in massive clusters drops slower and is higher at $r>1$ kpc, resulting in much shorter cooling times and stronger cooling flows in massive clusters. We will explore the role of gravity in the next subsection.

\begin{figure}
   \centering
\plotone {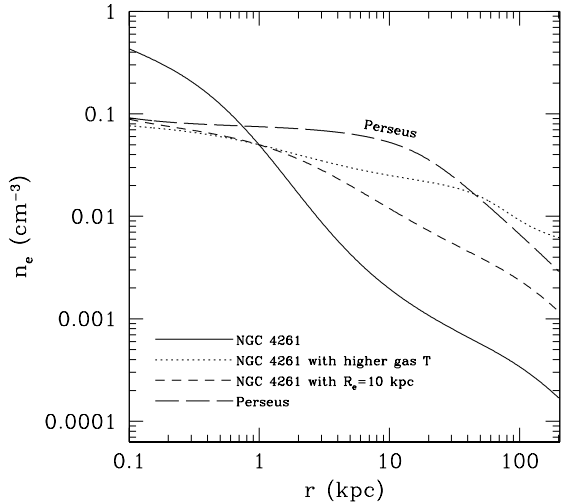} 
\caption{Initial gas density profiles in three simulations of NGC 4261 at $Z=0.3Z_{\sun}$. The solid line shows the profile in our standard simulation presented in Fig. \ref{plot4}. The dotted line shows the initial density profile when the initial gas temperatures are higher (using the initial temperature profile of Perseus; run NGC 4261-A1), while the short-dashed line is the initial density profile if its half-light radius is larger ($R_{e}=10$ kpc; run NGC 4261-A2). The long-dashed line shows the initial gas density profile of the massive cluster Perseus for comparison.}
 \label{plot11}
 \end{figure}

To test the role of gas temperature on the evolution of cooling flows, we performed an additional run (denoted as run NGC 4261-A1) for the group NGC 4261. In this run, we choose much higher initial gas temperatures by adopting the initial gas temperature profile from the Perseus cluster. From the initial gas temperature profile, we determine the initial gas density profile from hydrostatic equilibrium. We fixed the gas density at $r=1$ kpc to be the same as our standard run for NGC 4261, which is presented in the left panel of Fig. \ref{plot4}. The dotted line in Figure \ref{plot11} shows the resulting initial gas density profile in run NGC 4261-A1. As expected from equation (\ref{lrho}), the scale length of the density profile becomes larger and gas densities at large radii becomes much larger than in the standard run. The hot gas in this new run experiences a central cooling catastrophe at about $t=0.68$ Gyr. We note that in our standard run of this system, the hot intragroup gas does not experience a cooling catastrophe even at the end of the simulation ($t=3$ Gyr). This indicates that the low gas temperature and the associated low gas densities in galaxy groups and elliptical galaxies at large radii ($r \gtrsim 0.1$ - $1$ kpc) do play an important role in maintaining the long-term hot-mode transition from cooling flows to SMBH accretion flows.

   \begin{figure*}
   \centering
\plottwo {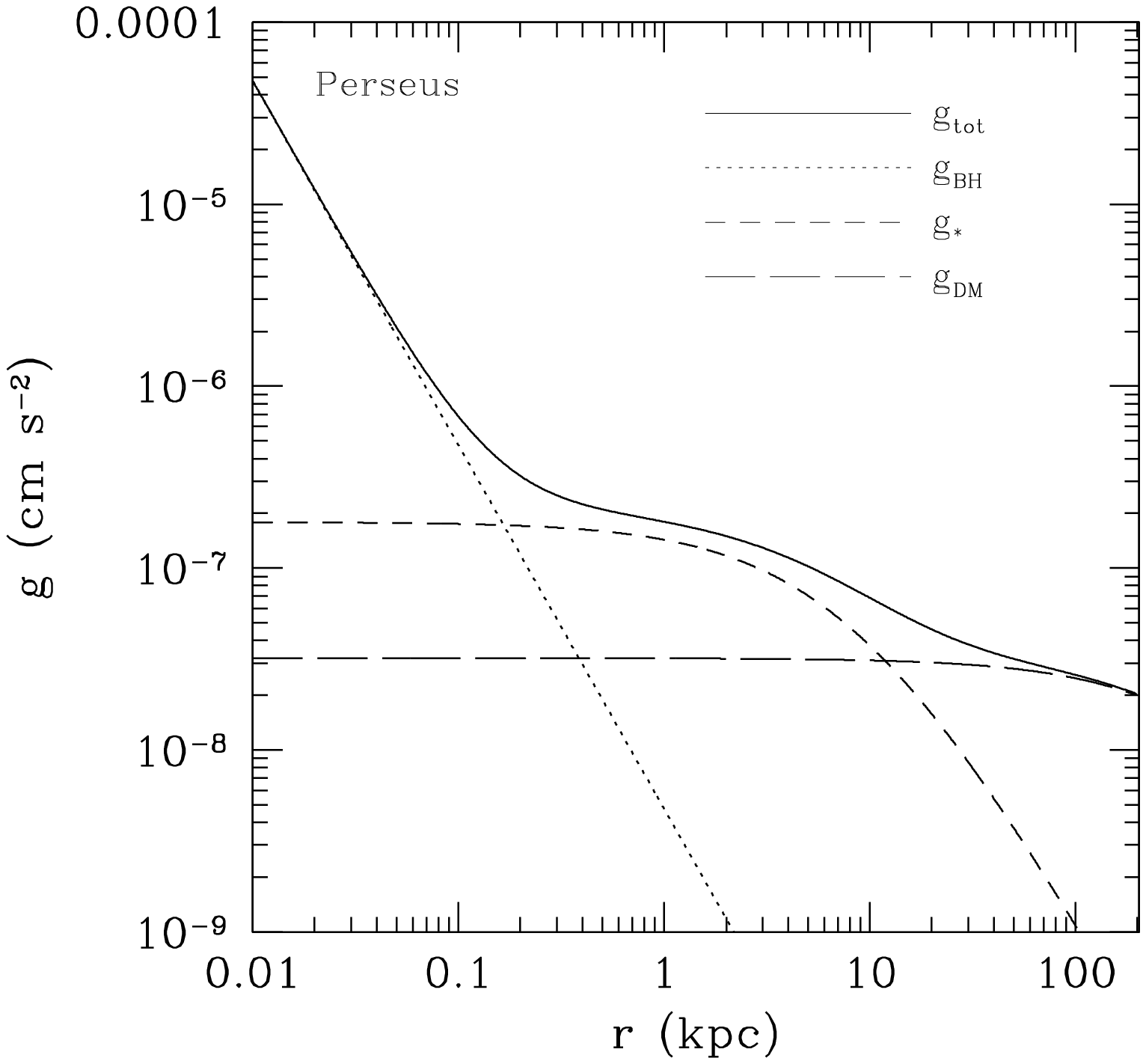} {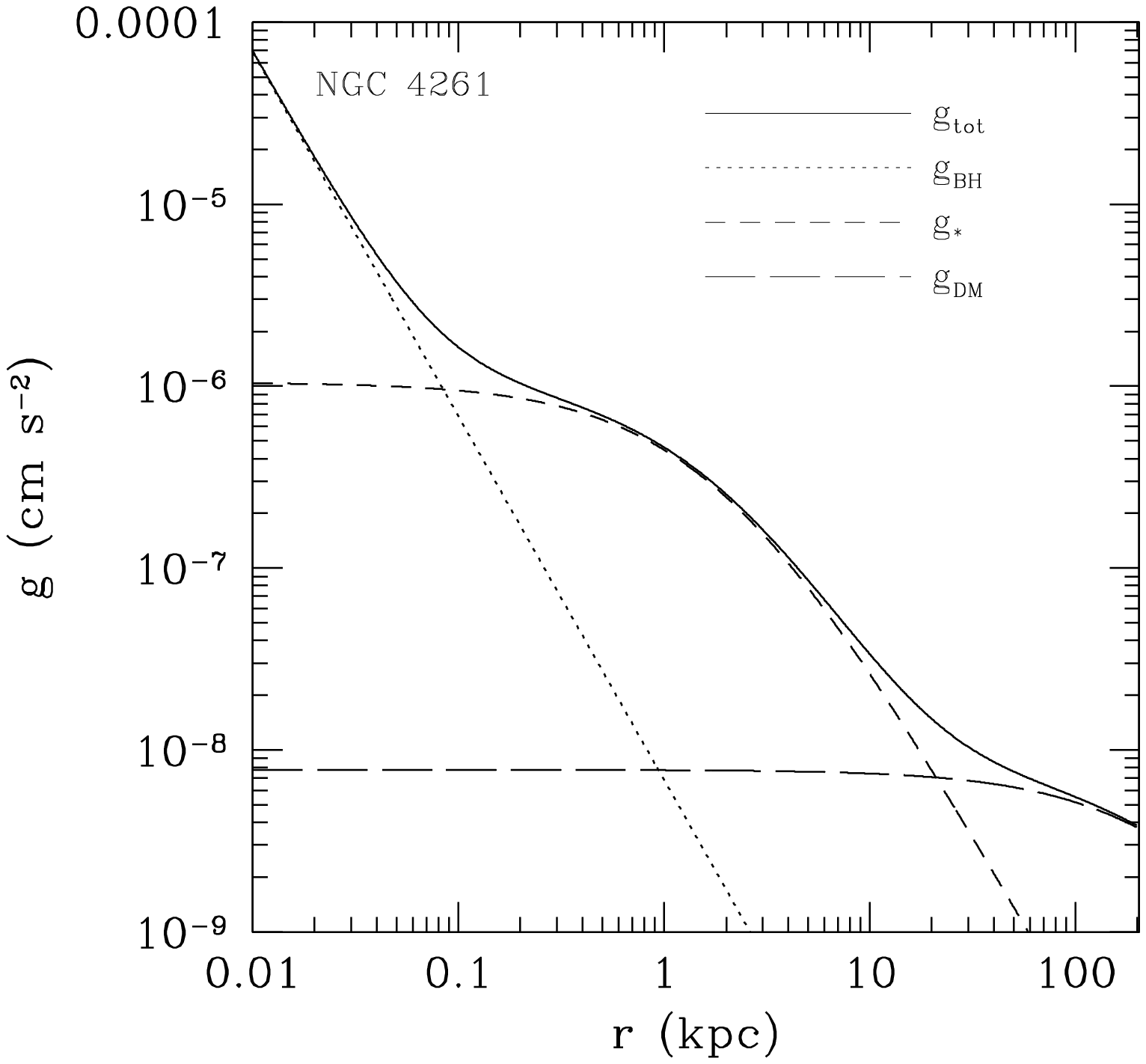}
\caption{The gravitational acceleration and its three contributing components in the massive cluster Perseus (left) and the galaxy group NGC 4261 (right). The gravity within $0.1 \lesssim r \lesssim 10$ kpc is dominated by the central galaxy, which is usually more compact (i.e., with a much smaller half-light radius $R_{e}$), and contributes much stronger gravity in smaller systems (galaxy groups and elliptical galaxies) than in massive clusters.}
 \label{plot12}
 \end{figure*} 
 
\subsection{The Gravity due to the Central Galaxy}

The role of gravity is two-fold. First, the gravitational acceleration affects the scale length of the equilibrium density profile. As seen in equation (\ref{lrho}), with a stronger gravitational acceleration $g=|\nabla \Phi|$, the scale length of the gas density profile is shorter, leading to smaller gas densities beyond about 1 kpc. Second, the gravity affects the development of the inflow velocity (the inflow timescale), which is important in determining the fate of cooling flows (eqs. \ref{tscale1} and \ref{tscale2}). The release of the gravitational energy in the cooling flow goes into the increase of the gas kinetic and thermal energy. This argument suggests that the gravitational acceleration may play a more important role than the gas temperature in the bimodal fate of cooling flows.

In Figure \ref{plot12}, we show the gravitational acceleration and its three contributing components in the massive cluster Perseus and the galaxy group NGC 4261. Here $g_{\rm BH}=|\nabla \Phi_{\rm BH}|$, $g_{*}=|\nabla \Phi_{*}|$, $g_{\rm DM}=|\nabla \Phi_{\rm DM}|$, and $g_{\rm tot}=g=g_{\rm BH}+g_{*}+g_{\rm DM}$. At $r\lesssim 0.1$ kpc, the gravity is dominated by the central SMBH, while at $r\gtrsim 10$ kpc, the gravity is dominated by the dark matter halo. In the middle radii $0.1 \lesssim r \lesssim 10$ kpc, where large-scale cooling flows are gradually converted to accretion flows of the central SMBH, the gravity is often dominated by the central galaxy. Comparing the left and right panels of Figure \ref{plot12}, it is clear that the gravitational acceleration at $0.1 \lesssim r \lesssim 10$ kpc (dominated by the central galaxy) is much larger in the group NGC 4261 (and other groups in our sample) than in the massive Perseus cluster (typically by a factor of more than 5). This is mainly due to the fact that the half-light radius $R_{e}$ of central galaxies in our group sample ($R_{e}\sim 3$ - $4$ kpc) is much smaller than that in our cluster sample (usually $R_{e} \gtrsim 10$ kpc). For the clusters Abell 1795 and Abell 2199, the half-light radius is very large ($\sim 40$ kpc; see Table 1), resulting in $g_{*}$ smaller than that in NGC 4261 by more than one order of magnitude. 

We tested if the size (gravity) of the central galaxy has a strong effect on the fate of cooling flows by performing an additional run (denoted as run NGC 4261-A2) for NGC 4261. In this run, we keep the initial gas temperature profile unchanged, but increase the half light radius from $3.4$ kpc to $10$ kpc. With a larger $R_{e}$, the gravitational acceleration contributed by the central galaxy becomes smaller and the initial equilibrium gas density profile becomes flatter, as seen in Figure \ref{plot11}. In this run, the intragroup gas experiences a cooling catastrophe very quickly at $t\sim 0.15$ Gyr. Compared to the standard simulation and run NGC 4261-A1, this confirms our expectation that the small size and the associated strong gravitational acceleration of the central galaxy play a key role in maintaining cooling flows of some galaxy groups and elliptical galaxies in hot-mode when transitioning into SMBH accretion flows.
 
We also performed a few numerical experiments based on the massive clusters Abell 1795 and Perseus. If we decrease the initial gas temperatures in these massive systems to about $1$ keV (a typical gas temperature in galaxy groups), the hot intracluster gas usually still suffers a central cooling catastrophe very quickly. But if we adopt a small size (e.g. $R_{e} \sim 4$ - $5$ kpc) for the central galaxy, the cooling catastrophe is usually significantly delayed. In these calculations, we also vary the initial gas density profile according to hydrostatic equilibrium as done in runs NGC 4261-A1 and NGC 4261-A2. This suggests that the size and gravity of the central galaxy are more important than gas temperature in producing the dichotomy in the transition of cooling flows to accretion flows discussed in Section \ref{section3}.

The fast increase of the half-light radius with stellar mass (or luminosity) is a general trend observed in early-type galaxies (e.g., \citealt{shen03}; \citealt{hyde09}). A large size of $R_{e}\sim 10$ - $40$ kpc is typical for central galaxies in galaxy clusters (e.g. \citealt{graham96}), and is also predicted in state-of-the-art cosmological simulations that include AGN feedback, which suppresses overcooling and star formation in cluster central regions and produces ``adiabatic expansion" of the stellar distribution through AGN-driven gas outflows \citep{martizzi12}. As discussed in this subsection, a larger size of the central galaxy corresponds to a relatively smaller gravitational acceleration at the radii $0.1 \lesssim r \lesssim 10$ kpc, which results in higher gas densities and stronger cooling flows at $r\gtrsim 1$ kpc, and less efficient gravitational heating on infalling cooling flows, leading to the quick development of a cooling catastrophe within about $100$ - $300$ Myr in massive clusters in our model.

Although central galaxies in galaxy clusters typically have large sizes ($R_{e}\gtrsim 10$ kpc), some may have slightly smaller sizes. The Virgo cluster is probably an extreme case --- its central galaxy (M87) has a very small half-light radius $R_{e}\sim 5$ kpc. Virgo is a small and dynamically young cluster \citep{binggeli87}, which may explain the small size of M87. The relatively low gas temperature, the small size of M87, and hence the strong gravitational acceleration from M87 explain why the cooling catastrophe in our Virgo simulation at  $Z=0.3Z_{\sun}$ happens at a relatively later time $t_{\rm cc}\sim 1.7$ Gyr (but the simulation at a higher metallicity $Z=Z_{\sun}$ results in a much smaller $t_{\rm cc}\sim 0.22$ Gyr). 

\section{Discussion and Observational Prospects}
\label{section5}

\subsection{Comparison with Previous Relevant Studies}

In this subsection, we briefly discuss a few previous works that are relevant to our study.

Spherically-symmetric accretion of cooling flows onto SMBHs has been studied by \citet{quataert00} and \citet{mathews12} with steady-state models, revealing the presence of both hot-mode and cold-mode solutions. But steady-state models usually do not predict which mode a real system chooses, which is investigated in our simulations of eight well-observed systems. Our results are consistent with numerical simulations of \citet{li12}, who found the cold mode accretion in one system -- Perseus. We further show that the hot-mode accretion of cooling flows can be present in some galaxy groups and elliptical galaxies.

One main feature of the central hot-mode accretion is its radially-decreasing temperature profile, which has also been seen in numerical simulations of \citet{brighenti02} and \citet{gaspari12}. However, these studies did not have the spatial resolution ($\sim 100$ pc) to accurately investigate the central accretion region, and adopted an ad-hoc mass dropout term for relatively cold gas, which still accumulates in central regions, differing significantly from the central hot-mode accretion in our studies.

Our idealized model explores the interplay between radiative cooling, gas inflow, and compressional heating in a spherically symmetric setup, while ignoring other potentially important physics. Recently \citet{gaspari13} studied the potential roles of turbulence and AGN heating on the accretion of cooling flows in a fiducial system -- NGC 5044. Their 3-dimensional (3D) simulations indicate that if the hot gas is maintained in a global thermodynamic balance by AGN heating, turbulence and cooling produce a chaotic cold accretion onto the central SMBH. 

\subsection{The Importance of Covering Both Cooling and Accretion Regions}

To properly study the cooling-flow-fed accretion onto SMBHs, it is important to cover both the cooling and accretion regions simultaneously in simulations. Partly due to the limitation of numerical resolution, previous simulations of AGN feedback often focus on kpc and larger scales with an inner boundary (or spatial resolution) of typically $1$ kpc. These simulations can not explore the interesting physics of the accretion of cooling flows near SMBHs on pc scales. We confirmed this by rerunning a few simulations covering a spatial range of $1$ - $200$ kpc. For galaxy groups, we got similar quasi-steady flows beyond 1 kpc, but the hot-mode accretion with negative temperature gradients are not seen as it is located within the inner boundary $r=1$ kpc. For galaxy clusters, the simulation results vary from system to system, probably depending on the SMBH mass and its associated compressional heating beyond 1 kpc. Abell 1795 (with a high SMBH mass) quickly reaches a cooling catastrophe near the inner boundary $r=1$ kpc, while our standard simulation covering $0.01$ - $200$ kpc (Figure \ref{plot1}) shows that the cooling catastrophe happens first at a concentric radius of tens pc and then quickly propagates inward and outward. For Abell 2199 and the Perseus cluster (with slightly smaller SMBH masses compared to Abell 1795), the simulations covering only $1$ - $200$ kpc did not reveal a cooling catastrophe, but instead reach a steady state with mass inflow rates of hundreds $M_{\sun}$/yr (similar to run A in \citealt{guo08a}).

More importantly, if only the central kpc or smaller scales are simulated, as often seen in accretion flow simulations, one can not study how cooling flows formed on kpc and larger scales feed the central SMBH, and the results depend significantly on the chosen outer boundary conditions. If the gas density and temperature are fixed at the outer boundary (assumed to be at ``infinity"), our simulation of Abell 1795 with a spatial range of $0.01$ - $1$ kpc results in a quasi-steady hot-mode accretion with central negative temperature gradients, strikingly different than the cooling catastrophe seen in our standard simulation for Abell 1795. This confirms that the central cooling catastrophe developed in cluster cooling flows is a global phenomenon: the cooling flow formed on kpc and larger scales (up to $\sim 100$ kpc in massive clusters) provides the gas inflow and sets the outer boundary conditions for the central accretion flow.

\subsection{Assumptions and Limitations of Our Model}

Assuming spherical symmetry, our 1D simulations study the development of cooling flows and the resulting accretion onto the central SMBHs under the competitive roles of radiative cooling and gravitational heating. We ignore other physical processes, which may also play significant roles during the flow evolution (as also discussed in Section 2).

For example, our 1D setup does not allow us to investigate the role of angular momentum in the cooling flow to accretion flow transition. X-ray observations of massive ellipticals, galaxy groups and clusters indicate that the level of angular momentum is usually not significant. Assuming a constant rotational velocity of $50$ km/s, as expected from cosmological cluster simulations, 3D simulations of the Perseus cluster by \citet{li12} showed that the cluster gas outside about $100$ pc can be well described by spherical symmetric flows. Nevertheless, angular momentum may play a significant role in central regions, in particular, for galaxy clusters when the gas cools off from the cooling catastrophe. The cold gas will likely form a torus and accretion disk, which are still hardly resolved by current 3D simulations of cooling flows \citep{li12}. For some smaller systems where the hot-mode accretion dominates in central regions, the role of angular momentum is less clear. \citet{narayan11} argued that viscosity in the hot gas, which increases strongly with the gas temperature, can effectively transport angular momentum outward of central regions. Investigating the role of angular momentum requires 3D simulations, but we expect that the bimodality in the cooling-flow-fed accretion is robust.

Another major assumption in our model is the neglect of AGN heating and turbulence. If not suppressed by thermal conduction, the perturbations of the hot gas by turbulence or AGN outbursts may produce the non-linear growth of local thermal instability, resulting in chaotic cold accretion onto the SMBHs as shown in \citet{gaspari13}. While the accreted cold gas in this mode may come from random directions, the cold gas from a global cooling catastrophe is expected to preserve the net angular momentum of the hot gas on large scales.

\subsection{Observational Prospects}

It is now commonly thought that cooling flows in galaxy groups and clusters are significantly suppressed and AGN feedback plays a key role in the suppression process. However, there may still be two evolution scenarios for the state of hot gas in cool core systems. If AGN feedback operates ``instantaneously" in response to cooling (e.g., \citealt{guo08a}), the hot gas may reach a quasi-steady cool-core state with cooling flows suppressed to a low level all the time. Recent studies of local thermal instability in cool core clusters are based on this scenario (\citealt{mccourt12}; \citealt{sharma12}; \citealt{gaspari12b}). The other scenario arises from the periodicity of AGN feedback events, assuming that the hot gas also evolves periodically. Between two successive generations of AGN feedback events, cooling flows may be established to some level, leading to the formation of cold gas and stars in the central galaxy and feeding the central SMBH. The triggered AGN outburst then heats the hot gas, strongly suppressing or shutting off the cooling flow. In a time-averaged sense, cooling may be balanced by AGN heating, similar to the first scenario. Here we briefly discuss observational prospects of our model in the second scenario.

Our model predicts that cold gas and star formation activities are present in central galaxies of massive clusters and some galaxy groups when the central cooling catastrophe develops, while some galaxy groups and elliptical galaxies are expected to have little cold gas and star formation activities due to the hot-mode accretion of cooling flows. For galaxy groups and ellipticals, the cold-mode accretion and cold gas prefer to happen in systems with high gas metallicities. 

Observations have detected significant star formation activities (H$\alpha$ emission and blue light) in BCGs, particularly in galaxy clusters with low central gas entropy or low central cooling times (\citealt{rafferty08}; \citealt{cavagnolo08}). This has been previously interpreted as an evidence for the development of local thermal instability (\citealt{mccourt12}; \citealt{sharma12}), while thermal conduction may also play a role in suppressing the instability in high-entropy systems \citep{voit08}. In our model, both star formation and low central entropy (cooling time) appear simultaneously in a much-developed cooling flow as the central cooling catastrophe starts. The estimated star formation rates are typically a few to tens $M_{\sun}$/yr \citep{odea08}, which is about $1/100$ to $1/10$ of the predicted gas inflow rate if the cooling catastrophe happens unimpeded. This is consistent with our picture that the cooling catastrophe, once starts and feeds the SMBH, is quickly suppressed or averted by the triggered AGN feedback event. Observations also suggest that star formation rates in elliptical galaxies are usually extremely small (e.g., $\sim 10^{-5}$ $M_{\sun}/$yr in \citealt{ford12}, about four orders of magnitude less than the steady-state mass inflow rate of $\sim0.1$ - $1$ $M_{\sun}/$yr in our simulations of four galaxy groups and ellipticals). 

Another observational consequence of the dichotomy is that central cusp temperature profiles (with negative gradients) are expected to be observed more often in elliptical galaxies and galaxy groups than in more massive galaxy clusters. In some galaxy groups and ellipticals, the central negative temperature gradient is maintained quasi-steadily in the hot-mode accretion unless AGN feedback events destroy it, while in galaxy clusters it quickly develops into a cooling catastrophe. Recent {\it Chandra} observations of the S0 galaxy NGC 3115 by \citet{wong11} found that the gas temperature within the central $\sim 200$ pc rises inward. Negative temperature gradients have also been seen in the group NGC 1407 \citep{humphrey06}, the elliptical galaxy NGC 4649 \citep{humphrey06} and NGC 4374 \citep{allen06}, usually covering a central region with spatial sizes ($r\lesssim 0.5$ - $1$ kpc) larger than those predicted in our calculations ($r\lesssim 50$ - $100$ pc). The rather large sizes suggest that they may be produced by other processes (e.g. a strong AGN outburst), but these central cuspy temperature profiles were all observed in galaxies or galaxy groups, which is consistent with our prediction. If some of them are indeed produced by gravitational heating, the mass of the central SMBH or the central stellar mass could be much larger than assumed in our calculations.

The two distinct SMBH accretion modes may trigger AGN feedback events with different observational features. One difference is the SMBH accretion rate, which is about $\sim0.1$ - $1$ $M_{\sun}/$yr in the hot-mode accretion of our simulated galaxy groups and ellipticals, while in our four simulated galaxy clusters, it increases quickly to several hundreds $M_{\sun}/$yr after the cooling catastrophe starts. The significantly larger central mass inflow rates in massive clusters are essentially determined by the larger cooling flow rates formed on kpc and larger scales in these systems, and naturally explain why AGN feedback events in clusters are typically much more powerful than in smaller systems. Other than energetics, the feedback events triggered by the hot- and cold-mode accretions may be intrinsically different, though the SMBH spin may also play a significant role. Further theoretical and observational studies, in particular focusing on the difference between these two modes, are required to explore this issue.

 \section{Summary}
\label{section:conclusion}

The importance of AGN feedback on the evolution of elliptical galaxies, galaxy groups and clusters has received great appreciation, in particular, during the last decade. It is widely thought that AGN feedback plays a key role in suppressing strong cooling flows predicted in standard cooling flow models. Here we perform a series of hydrodynamic simulations for many systems, ranging from elliptical galaxies to massive galaxy clusters. The main purpose is to answer one of the key questions in AGN feedback: How are cooling flows accreted by central SMBHs? We aim to explore the triggering mechanism of AGN feedback, providing an important link between cooling flows and their triggered AGN feedback events which eventually avert the further development of cooling flows.

Our simulations follow the evolution of cooling flows on scales as large as about $200$ kpc down to SMBH accretion flows on pc scales in eight well-observed systems. The development of cooling flows on large scales affects mass inflow rates and outer boundary conditions of SMBH accretion flows on small scales. By covering such a large spatial range, we are able to self-consistently study the cooling-flow-induced accretion onto the SMBHs. For initial conditions, we adopt the gas temperature and density profiles derived from previous X-ray observations. In our model, the gravitational potential well is contributed by a SMBH, a central galaxy and the dark matter halo, and the relevant parameters are all based on the estimates from available multi-wavelength data. 

Our calculations reveal an interesting dichotomy in the transition of cooling flows to accretion flows. In galaxy clusters, a global cooling catastrophe develops in central regions roughly within the gas cooling time (typically $100$ - $300$ Myr), resulting in a cold-mode accretion onto the central SMBH. During the cooling catastrophe, the mass inflow rate increases dramatically and quickly to about several hundreds $M_{\sun}/$yr, close to the values predicted in the standard cooling flow model. However, the development of cooling flows is very different in some smaller systems, including galaxy groups and elliptical galaxies. Despite of short central cooling times, the hot gas in these systems quickly settles into a quasi-steady state, which lasts for several Gyr or even permanently. In this state, the gas temperature in central regions (typically within tens to about 100 pc, depending on the SMBH mass) rises toward the center, indicating a hot-mode accretion onto the SMBH. The hot-mode accretion is present quasi-steadily in all of our four small systems if adopting a gas metallicity $Z=0.3Z_{\sun}$, and in half of them if adopting a more realistic metallicity $Z=Z_{\sun}$.

Both accretion modes naturally appear in our idealized simulations where only radiative cooling, spherically-symmetric gas inflow, and compressional heating are taken into account. These three physical processes are entangled and determine the development of cooling flows. In galaxy clusters, the cooling time is significantly shorter than the inflow time and the compressional heating time across almost the whole cooling regions (except the very central regions where the compressional heating due to the SMBH's gravity dominates), and the hot gas cools unimpeded, resulting in a central cooling catastrophe at about the cooling time. However, in some galaxy groups and elliptical galaxies, the approximate balance between these three timescales is quickly built up from small to large radii, leading to a quasi-steady state with a typical mass inflow rate of about $\sim0.1$ - $1$ $M_{\sun}/$yr. 

Our calculations indicate that the long-lasting hot-mode accretion of cooling flows prefers to happen in elliptical galaxies and galaxy groups with low gas densities, low gas metallicities, and importantly, compact central galaxies, which result in strong gravitational acceleration and compressional heating in the transitioning regions from cooling flows to accretion flows (about $0.1$ - $10$ kpc). These requirements may not be needed simultaneously. For example, two groups, NGC 4261 and NGC 1407 (with $R_{e}=3.4$ and $4.4$ kpc respectively), host the hot-mode accretion in our calculations even at the solar metallicity. The four small systems in our sample may be biased toward those with compact central galaxies, and the real fraction of groups hosting the long-term hot-mode accretion of cooling flows may be much lower. A more systematic study is required to explore the general behavior of cooling flows in galaxy groups.

It is important to note that our 1D simulations ignore angular momentum, AGN heating, turbulence, transport processes, and stellar mass losses from the central galaxy. These additional processes may affect the accretion flow, and if significant, even change the accretion mode, as discussed in Section 5.

Our calculations predict that cuspy temperature profiles are observed more often in elliptical galaxies and galaxy groups, while massive clusters often contain more cold gas and stronger star formation activities in their central galaxies. Recent observations indeed found that many BCGs contain large amounts of cold gas and significant star formation activities. Future X-ray observations with high spatial resolutions will be very useful in exploring the gas properties and the interesting plasma physics within the SMBH's gravitational influence.

\acknowledgements 

FG acknowledges the support by the Zwicky Prize Fellowship of ETH Z\"{u}rich and thanks the generous hospitality of Department of Astronomy at University of Science and Technology of China, where part of the work was done. We thank the referee for insightful comments, which lead us to explore the dependence of our results on the gas metallicity, significantly improving the work. Studies of black hole growth and feedback at UC Santa Cruz are supported by NSF and NASA grants for which we are very grateful.


\begin{appendix}
\section{Dependence on the Choices of the Inner and Outer Boundaries}
  \begin{figure}
   \centering
\plotone {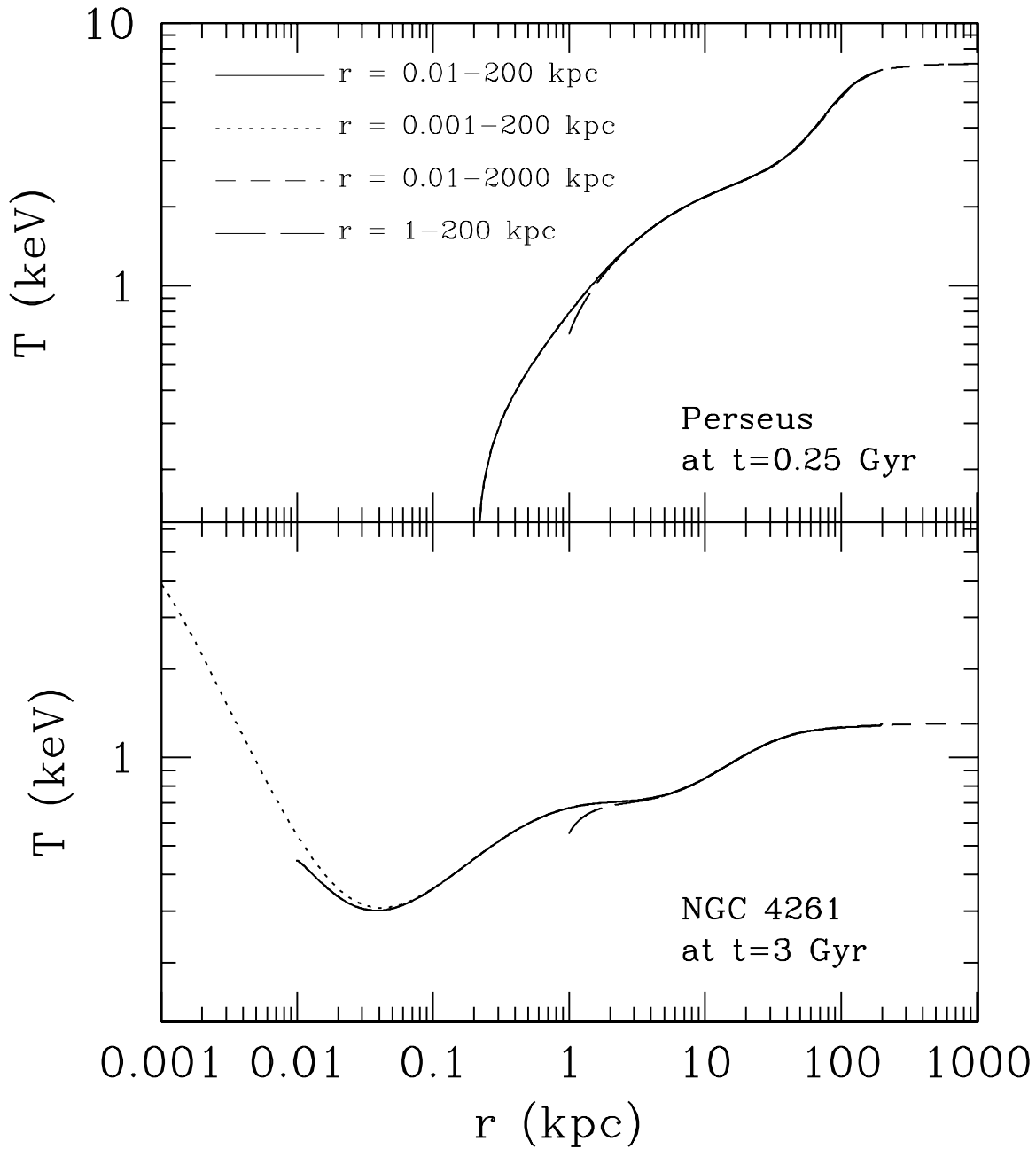} 
\caption{The gas temperature profiles in a series of four simulations with different boundary locations for Perseus at $t=0.25$ Gyr (top) and for NGC 4261 at $t=3$ Gyr (bottom), respectively. The four simulations for each system cover a spatial range of $10$ pc - $200$ kpc (solid line),  $1$ pc - $200$ kpc (dotted line),  $10$ pc - $2$ Mpc (short-dashed line) and $1$ - $200$ kpc (long-dashed line), respectively.}
 \label{plot13}
 \end{figure} 
 
Our standard simulations presented in Sections 3.1 and 3.2 cover a spatial range from $r_{\rm min}=10$ pc to $r_{\rm max}=200$ kpc, which allows to explore both the cooling flow and accretion flow regions. To investigate how our results depend on the choices of the inner and outer boundaries, we have also performed a few additional simulations, some of which are presented in this Appendix. 

Figure \ref{plot13} shows the gas temperature profiles in a series of four simulations with different boundary locations for Perseus at $t=0.25$ Gyr (top) and for NGC 4261 at $t=3$ Gyr (bottom). The four simulations for each system cover a spatial range of $10$ pc - $200$ kpc (solid line; the standard run in the paper),  $1$ pc - $200$ kpc (dotted line),  $10$ pc - $2$ Mpc (short-dashed line) and $1$ kpc - $200$ kpc (long-dashed line), respectively. As clearly seen in the Figure, changing the outer boundary from 200 kpc to 2Mpc has a negligible impact on the final temperature profile for both systems, mainly due to the fact that the simulation time is much shorter than the gas cooling time at these large outer boundaries.

For the inner boundary, we experimented with three values $r_{\rm min}=1$, $10$, and $1000$ pc. As shown in Figure \ref{plot13}, all simulations capture the gas temperature evolution quite well in their simulated regions except near the inner boundary (probably due to the inner boundary conditions). However, the simulation with $r_{\rm min}=1$ kpc can not capture the interesting physics happening at $r<r_{\rm min}$, including the central cooling catastrophe in Perseus and the central cuspy temperature profile in NGC 4261. On the other hand, the difference between the simulations with $r_{\rm min}=1$ and $10$ pc is small. For Perseus, both simulations show the central cooling catastrophe with almost the same size at $t=0.25$ Gyr. For NGC 4261, both simulations reveal the central region with negative temperature gradients, though the simulation with $r_{\rm min}=1$ pc shows a steeper temperature slope there, indicating slightly stronger compressional heating.

\end{appendix}

\label{lastpage}

\end{document}